\documentclass[
  aps,
  pra,
  reprint,
  superscriptaddress,
  amsmath,
  amssymb,
  amssymb,
  amsfonts,
]{revtex4-2}

\usepackage{dcolumn}
\usepackage{bm}
\usepackage{hyperref}
\usepackage{xcolor}
\usepackage{graphicx}
\graphicspath{{./figures/}}
\usepackage{tikz}
\usepackage{physics}
\usepackage{pgfplots}
\pgfplotsset{compat=1.18}
\usepackage{ifthen}
\usetikzlibrary{positioning, calc}
\usetikzlibrary{decorations.pathmorphing} 
\usetikzlibrary{arrows.meta} 
\usepackage[export]{adjustbox}
\usepackage{xspace} 
\usepackage{pgf}
\usepackage{bbold}
\usepackage{soul}
\usepackage{lmodern}
\usepackage{placeins}
\usepackage{mathrsfs}
\usepackage[inline]{enumitem}
\usepackage{booktabs}
\usepackage{float}
\usepackage{multirow}

\newcommand{\mathdefault}[1][]{} 
\newcommand{\avg}[1]{\langle {#1} \rangle}

\newcommand{\red}[1]{{\color[rgb]{0.7,0,0} #1}}

\setstcolor{red}
\makeatletter
\newcommand{\hrules}{%
\@for\i:=1,2,3,4,5\do{%
\noindent\red{\rule{\linewidth}{1pt}}\par\vspace{-1ex}%
}%
\vspace{1ex}
}
\makeatother

\newcommand{\Htot}{H_\mathrm{tot}}
\newcommand{\Hctrl}{H_\mathrm{ctrl}}
\newcommand{\Hs}{H_\mathrm{S}}
\newcommand{\Hb}{H_\mathrm{B}}
\newcommand{\Hsb}{H_\mathrm{SB}}
\newcommand{\Hi}{H_\mathrm{I}}

\begin{document}
  \setlength{\columnsep}{20pt}

  \title{Machine Learning-aided Optimal Control of a noisy qubit}



  \author{Riccardo Cantone}
  \affiliation{Dipartimento di Fisica e Astronomia ``Ettore Majorana'', Università di Catania, Via S. Sofia 64, 95123 Catania, Italy}

  \author{Shreyasi Mukherjee}
  \affiliation{Dipartimento di Fisica e Astronomia ``Ettore Majorana'', Università di Catania, Via S. Sofia 64, 95123 Catania, Italy}

  \author{Luigi Giannelli}
  \affiliation{Dipartimento di Fisica e Astronomia ``Ettore Majorana'', Università di Catania, Via S. Sofia 64, 95123 Catania, Italy}
  \affiliation{Istituto Nazionale di Fisica Nucleare, Sezione di Catania, 95123, Catania, Italy}

  \author{Elisabetta Paladino}
  \affiliation{Dipartimento di Fisica e Astronomia ``Ettore Majorana'', Università di Catania, Via S. Sofia 64, 95123 Catania, Italy}
  \affiliation{Istituto Nazionale di Fisica Nucleare, Sezione di Catania, 95123, Catania, Italy}
  \affiliation{CNR-IMM, Via S. Sofia 64, 95123, Catania, Italy}

  \author{Giuseppe Falci}
  \affiliation{Dipartimento di Fisica e Astronomia ``Ettore Majorana'', Università di Catania, Via S. Sofia 64, 95123 Catania, Italy}
  \affiliation{Istituto Nazionale di Fisica Nucleare, Sezione di Catania, 95123, Catania, Italy}
  \affiliation{CNR-IMM, Via S. Sofia 64, 95123, Catania, Italy}



  \date{\today}

  \begin{abstract}
    We apply a graybox machine-learning framework to model and control a qubit undergoing
    Markovian and non-Markovian dynamics from environmental noise. The approach
    combines physics-informed equations with a lightweight transformer neural
    network based on the self-attention mechanism. The model is trained on simulated
    data and learns an effective operator that predicts observables accurately,
    even in the presence of memory effects. We benchmark both non-Gaussian random-telegraph
    noise and Gaussian Ornstein-Uhlenbeck noise and achieve low prediction errors
    even in challenging noise coupling regimes. Using the model as a dynamics
    emulator, we perform gradient-based optimal control to identify pulse sequences
    implementing a universal set of single-qubit gates, achieving fidelities
    above 99\% for the lowest considered value of the coupling and remaining above
    90\% for the highest.
  \end{abstract}


  \maketitle



  \section{Introduction}
  With the rapid development of quantum technologies, quantum control has emerged
  as a fundamental tool in enabling advancements in quantum computation,
  communication and sensing~\cite{ball2021software,abbott2020communication,poggiali2018optimal,
  soare2014experimental}. The goal of quantum control is to find carefully tailored
  external fields to manipulate efficiently a quantum system, for instance, to steer
  it toward a desired target state or operation~\cite{dong2010quantum}. Over the
  years, a variety of control strategies such as quantum optimal control~\cite{giannelli2022tutorial},
  feedback control~\cite{mabuchi2008coherent, clark2016quantum,
  karmakar2025noise}, dynamical decoupling~\cite{DD1, DD3, DD5} and adiabatic control~\cite{turyansky2025pulse,
  zeng2019adiabatic, mukherjee2024noise} have been developed to serve purposes ranging
  from suppressing environmental noise, accelerating quantum dynamics, achieving
  high-fidelity state transfer or quantum gates.

  Achieving robust control in realistic open quantum systems remains a significant
  challenge, particularly in the presence of complex noise environments~\cite{delben2023control,
  ortega2024unifying}. Mathematically, quantum control requires minimization of
  a cost function with respect to a set of control parameters, the cost function
  itself depending strongly on the underlying system and noise model. Therefore,
  designing effective control pulses often relies on the knowledge of both the Hamiltonian
  principal system and of details of its interaction with the environment. This
  latter information is not easily accessible, especially when the environment presents
  non-Markovian and strongly coupled degrees of freedom. To overcome these
  limitations, Youssry et al.~\cite{youssri1,auza,youssri2, youssri3} proposed a
  \textit{graybox approach}, that integrates model-based methods accounting for
  the dynamics of the principal system with a machine learning (ML) approach
  extracting information on the environment from patterns of data and not from
  the explicit knowledge of the model. The aim is to construct a control framework
  that is both assumption-free and scalable to more complex quantum systems.

  Building on this idea, in this work, we develop an enhanced greybox model for emulating
  the system evolution, which is specialized in order to perform optimal control.

  The greybox model consists of two components: a \textit{whitebox part}, which computes
  the portion of the system dynamics that can be evaluated using known physics and
  analytically tractable expressions, and a \textit{blackbox part}, implemented
  via neural networks~\cite{gurney2018introduction, marquardt2021machine}, which
  learns the remaining contributions that are inaccessible or intractable through
  physical modelling. This hybrid structure allows us to retain physical insight
  while leveraging data-driven learning to handle the unknown or hardly
  tractable aspects of the open system dynamics.

  We first train the enhanced greybox model to optimally emulate the evolution
  of a driven single qubit undergoing pure dephasing. We consider two distinct environmental
  noise scenarios: (i) Random Telegraph Noise (RTN) and (ii) Ornstein-Uhlenbeck
  (OU) noise. In this work, we train the neural network by synthetic data
  obtained by simulating the qubit dynamics. A dataset is generated consisting of
  input-output pairs, where the inputs are control parameters, e.g., describing the
  external driving fields, and the outputs are the fidelities of a universal set
  of single-qubit gates. Using this supervised learning approach~\cite{geron2022hands,
  nasteski2017overview}, the black box component of the greybox model is trained
  to infer parameters that encapsulate the effects of the environment. By combining
  the outputs of both the whitebox (physics-informed) and blackbox (data-driven)
  components, we are able to accurately emulate the system's dynamics across a
  broad range of system-environment coupling strengths.

  After the training is complete, we employ a gradient-based optimal control technique
  to design control pulses that implement the universal set of single-qubit gates.
  For the lowest value of the coupling we considered, we achieve gate fidelities
  exceeding $99\%$, and even for the intermediate and highest values, the fidelities
  remain above $90\%$.

  The article is organized as follows. In Sec.~\ref{sec:system} we introduce the
  system Hamiltonian along with the noise models considered. Additionally, we also
  introduce some important quantities useful for the construction of the greybox
  model. In Sec.~\ref{sec:machine_learning}, we describe the greybox machine
  learning framework in detail. In ~\ref{sec:results}, we present results for the
  case of RTN and OU noise. In both scenarios, we outline the training methodology
  and demonstrate how the model is used for optimal quantum control. Finally, we
  conclude in Sec.~\ref{sec:conclusions} with a summary of our findings and
  discuss future directions, including scaling the approach to multiqubit systems
  and more complex noise environments.

  \section{\label{sec:system}System and Model}
  \subsection{\label{sec:hamiltonian}Hamiltonian}
  We consider a single qubit subject to classical dephasing noise along the $z$-axis
  and driven by external control fields. In the interaction picture, the dynamics
  is described by the time-dependent Hamiltonian (see Appendix~\ref{app:hamderivation}
  for details)
  \begin{equation}
    H(t) = \Hctrl(t) + g\beta(t)\,\sigma_{z}, \label{totalham}
  \end{equation}
  where $g$ is the coupling strength between the qubit and the noise, and
  $\beta( t)$ is a classical stochastic process modeling dephasing noise
  \cite{auza}. Specifically, in this work we consider $\beta(t)$ to be either a
  Random Telegraph Noise (RTN) process or an Ornstein-Uhlenbeck (OU) process.

  The control Hamiltonian $\Hctrl(t)$ implements a drive along the $x$ and $y$-axes
  \begin{equation}
    \label{eq:cntrlham}\Hctrl(t) = f_{x}(t)\sigma_{x}+ f_{y}(t)\sigma_{y},
  \end{equation}
  where each control field $f_{\alpha}(t)$, with $\alpha \in \{x, y\}$, consists
  of $N$ Gaussian-shaped pulses
  \begin{equation}
    \label{eq:controls}f_{\alpha}(t) = \sum_{k=1}^{N}A_{k,\alpha}\,\exp\left[-\frac{(t
    - \tau_{k})^{2}}{\sigma^{2}}\right].
  \end{equation}
  Here $A_{k,\alpha}$ is the amplitude of the $k$-th pulse along the $\alpha$-axis,
  $\tau_{k}= \frac{k}{N + 1}T$ with $i=1,\dots, N$ specifies the temporal position
  of the pulse over the total duration $T$, and $\sigma = \frac{T}{12N}$ is the
  pulse width. This choice of $\sigma$ ensures minimal overlap between adjacent pulses
  \cite{qdataset}. For our analysis we chose $N=5$.

  \subsection{\label{sec:noise}Noise}
  We consider two types of classical stochastic processes to model the noise
  acting on the qubit: Random Telegraph Noise~\cite{gardiner, mandel-wolf,
  papoulis} and the Ornstein-Uhlenbeck process~\cite{gardiner, mandel-wolf,
  papoulis, OU1, OU2}.

  \paragraph{Random Telegraph Noise.}
  RTN is a non-Gaussian, discrete-state stochastic process that randomly switches
  between two values, $+1$ and $-1$, with a switching rate $\gamma$. We denote
  the RTN process by $\beta_{\text{RTN}}(t)$, defined as:
  \begin{equation}
    \beta_{\text{RTN}}(t) = \beta_{\text{RTN}}(0) (-1)^{n(0,t)},
  \end{equation}
  where $\beta_{\text{RNT}}(0)$ is chosen uniformly at random from $\{+1, -1\}$,
  and $n(0,t)$ is the number of flips that occur in the time interval $[0,t]$,
  distributed according to a Poisson process with mean $\gamma t$. The two-time correlation
  function of RTN is given by
  \begin{equation}
    \langle \beta_{\text{RTN}}(t_{1}) \beta_{\text{RTN}}(t_{2}) \rangle = e^{-2\gamma
    |t_1 - t_2|}.
  \end{equation}

  RTN is particularly relevant for modeling bistable fluctuators, such as background
  charges in superconducting qubits based on Josephson junctions \cite{falci1, falci2}.
  Moreover, ensembles of RTN processes with different switching rates can reproduce
  $1/f$ noise spectra, a hallmark of solid-state quantum devices
  \cite{paladino2, supercondqubit}. Since RTN is non-Gaussian, its dynamics
  cannot be fully captured by the power spectrum alone; higher-order statistical
  moments significantly influence the system's evolution.

  \paragraph{Ornstein-Uhlenbeck Noise.}
  The OU process is a continuous-time, Gaussian, mean-reverting stochastic
  process widely used to model temporally correlated noise. Its steady-state two-time
  correlation function is
  \begin{equation}
    \langle \beta_{\text{OU}}(t) \beta_{\text{OU}}(s) \rangle = \frac{D}{2k}e^{-k
    |t-s|},
  \end{equation}
  where $k$ is the relaxation rate and $D$ controls the noise strength. In our
  analysis the parameters $D$ and $k$ are chosen so that the power spectrum of
  the OU process matches that of the RTN.

  Unlike RTN, OU noise is Gaussian, meaning it is fully characterized by its
  mean and autocorrelation function. This distinction is crucial: even when RTN
  and OU are engineered to share the same power spectral density, their different
  higher-order statistical properties lead to different qubit responses under control
  protocols. In particular, non-Gaussianity introduces additional challenges for
  dynamical decoupling and noise spectroscopy, motivating the development of
  tailored control strategies~\cite{falci3}.

  \subsection{\label{Methodology}Methodology}
  In this subsection, we briefly report the derivation of an expression for the expectation
  value of a generic operator $O$, formulated in a way that is compatible with
  the neural network framework presented in the next section. The full derivation
  can be found in~\cite{youssri1}. Starting from the total Hamiltonian in Eq.~\eqref{totalham},
  the joint state of system and bath evolves as
  \begin{align}
    \rho(t) & \;=\; U(t)\,\rho(0)\,U^{\dagger}(t),          \\
    U(t)    & \;=\; \mathcal{T}_{+}e^{-i\int_0^t H(s)\,ds}.
  \end{align}
  We seek the noise-averaged expectation of a qubit observable $O$ at the final time
  $T$,
  \begin{equation}
    \label{standard_exp}\mathbb{E}[O(T)]_{\rho}= \Bigl\langle \tr_{S,B}\bigl[\,U(
    T)\,(\rho(0)\otimes\rho_{B})\,U^{\dagger}(T)\,O\bigr]\Bigr\rangle,
  \end{equation}
  where $\rho_{B}$ is the bath's initial state and $\langle\cdot\rangle$ denotes
  averaging over the classical noise $\beta(t)$. It is possible to show that Eq.~(\ref{standard_exp})
  can be written as
  \begin{equation*}
    \mathbb{E}[O(T)]_{\rho}= \text{tr}_{S}\Big[ \big\langle O^{-1}\tilde{U}_{I}^{\dagger}
    (t)O\tilde{U}_{I}(t) \big\rangle U_{\text{ctrl}}(t)\rho(0)U_{\text{ctrl}}^{\dagger}
    (t)O \Big],
  \end{equation*}
  where $U_{ctrl}(t)$ is the time-ordered dynamical evolution unitary tied to
  $H_{ctrl}$ and $\tilde U_{I}$ captures all noise effects in a convenient
  rotating frame \cite{youssri1}.

  We therefore define the $V_{O}$ operator
  \begin{align}
    V_{O} & \;=\; \big\langle O^{-1}\tilde{U}_{I}^{\dagger}(t)O\tilde{U}_{I}(t) \big\rangle, \label{eq:VO_def}\intertext{so that}\mathbb{E}[O(T)]_{\rho} & \;=\; \text{tr}_{S}\big[ V_{O}U_{\text{ctrl}}(t)\rho(0)U_{\text{ctrl}}^{\dagger}(t)O \big]\label{expect}.
  \end{align}

  All stochastic and system-bath couplings are captured by $V_{O}$, while the
  other operators in eq.~\eqref{expect} depend only on controls. The operator $V_{O}$
  encapsulates all the information about the noise, its interaction with the
  control fields, and its effects on the system dynamics. Notably, this form is independent
  of the qubit's initial quantum state, making it broadly applicable across
  various scenarios. It can be shown~\cite{youssri1} that for purely classical noise
  and a traceless observable $O$, the operator $V_{O}$ can be written as
  \begin{equation}
    V_{O}= O^{-1}\,Q\,D\,Q^{\dagger}, \label{VO}
  \end{equation}
  with
  \begin{align*}
    D & = \begin{pmatrix}\mu&0\\0&-\mu\end{pmatrix},                                                                                                                                                 \\
    Q & = \begin{pmatrix}e^{i\psi}&0\\0&e^{-i\psi}\end{pmatrix} \begin{pmatrix}\cos\theta&\sin\theta\\-\sin\theta&\cos\theta\end{pmatrix} \begin{pmatrix}e^{i\Delta}&0\\0&e^{-i\Delta}\end{pmatrix}.
  \end{align*}
  Importantly, $V_{O}$ depends only on four real parameters,
  $\mu, \theta, \psi, \Delta$ (with $\mu\in [0,1]$).

  \subsection{\label{sec:optimalcontrol}Optimal Control}
  In our work, we use the $V_{O}$ operator within a machine learning architecture
  to compute a tomographically complete set of expectation values, which are
  then used to reconstruct the process matrix associated with the noisy system
  evolution~\cite{tensornet}. From this process matrix, we evaluate the process
  fidelity with respect to a universal set of single-qubit gates. In particular,
  we consider the following gates:
  \begin{equation}
    \label{eq:Qgates}
    \begin{aligned}
      I = \begin{pmatrix}1 & 0 \\ 0 & 1\end{pmatrix}, \quad X = \begin{pmatrix}0 & 1 \\ 1 & 0\end{pmatrix}, \quad Y = \begin{pmatrix}0 & -i \\ i & 0\end{pmatrix},                                              \\
      Z = \begin{pmatrix}1 & 0 \\ 0 & -1\end{pmatrix}, \quad H = \frac{1}{\sqrt{2}}\begin{pmatrix}1 & 1 \\ 1 & -1\end{pmatrix},                                                                                 \\
      R_{X}\left(\frac{\pi}{4}\right) = \begin{pmatrix}\cos\left(\frac{\pi}{8}\right) & -i\,\sin\left(\frac{\pi}{8}\right) \\ -i\,\sin\left(\frac{\pi}{8}\right) & \cos\left(\frac{\pi}{8}\right)\end{pmatrix}.
    \end{aligned}
  \end{equation}
  Once fully trained on the appropriate data, the machine learning model serves
  as a fast and computationally efficient emulator of the quantum system
  specialized for the optimization of the control pulses for implementing the
  quantum gates in eq.~\eqref{eq:Qgates}. Its function is to compute process matrix
  fidelities starting from a given set of control pulses.

  We use it to implement a quantum optimal control method aimed at finding sequences
  of Gaussian control pulses of the type of eq.~\eqref{eq:controls} that best realize
  the target set of gates~\cite{giannelli,wilhelm,oc2}.

  The \textit{cost functional} that we use as a quantitative measure of the
  deviation between the actual and desired system behavior is
  \begin{equation}
    \label{infidelity}J(u, \Theta; G) = 1 - \mathcal{F}(u, \Theta; G),
  \end{equation}
  where $\mathcal{F}(u, \Theta; G)$ is the \textit{process matrix fidelity},
  defined as
  \begin{equation}
    \label{eq:processmatrixfidelity}
    \begin{aligned}
      \mathcal{F}(u, \Theta; G) & = \mathcal{F}(\chi_{\text{actual}}(u, \Theta), \chi_{\text{target}}(G))                        \\
                                & = \text{tr}\!\left(\chi^{\dagger}_{\text{actual}}(u, \Theta)\, \chi_{\text{target}}(G)\right).
    \end{aligned}
  \end{equation}
  Here, $\chi_{\text{actual}}(u, \Theta)$ denotes the process matrix obtained as
  output of the machine learning model, while $\chi_{\text{target}}(G)$
  corresponds to the desired target process matrix~\cite{auza,tensornet}. The process
  matrix fidelity $\mathcal{F}$ depends on the pulse parameters $u(t)$, the
  fixed weights $\Theta$ of the trained machine learning model, and the target gate
  $G$.

  In order find the pulses that best give the wanted target gate, we minimize
  $J( u, \Theta; G)$ with a variant of the Broyden-Fletcher-Goldfarb-Shanno (BFGS)
  algorithm~\cite{optimizers}.

  \section{\label{sec:machine_learning}The machine learning model}
  We employ a hybrid machine learning architecture that integrates physics-informed
  analytical equations (whitebox) with neural networks (blackbox). This graybox paradigm~\cite{youssri1}
  uses the whitebox component to enforce the known laws of quantum mechanics,
  while the neural networks learn the effect of noise on the system.

\begin{figure}[h]
    \centering
    \includegraphics[width=0.4\textwidth, trim=0.1cm 0cm 0.2cm 0cm, clip]{
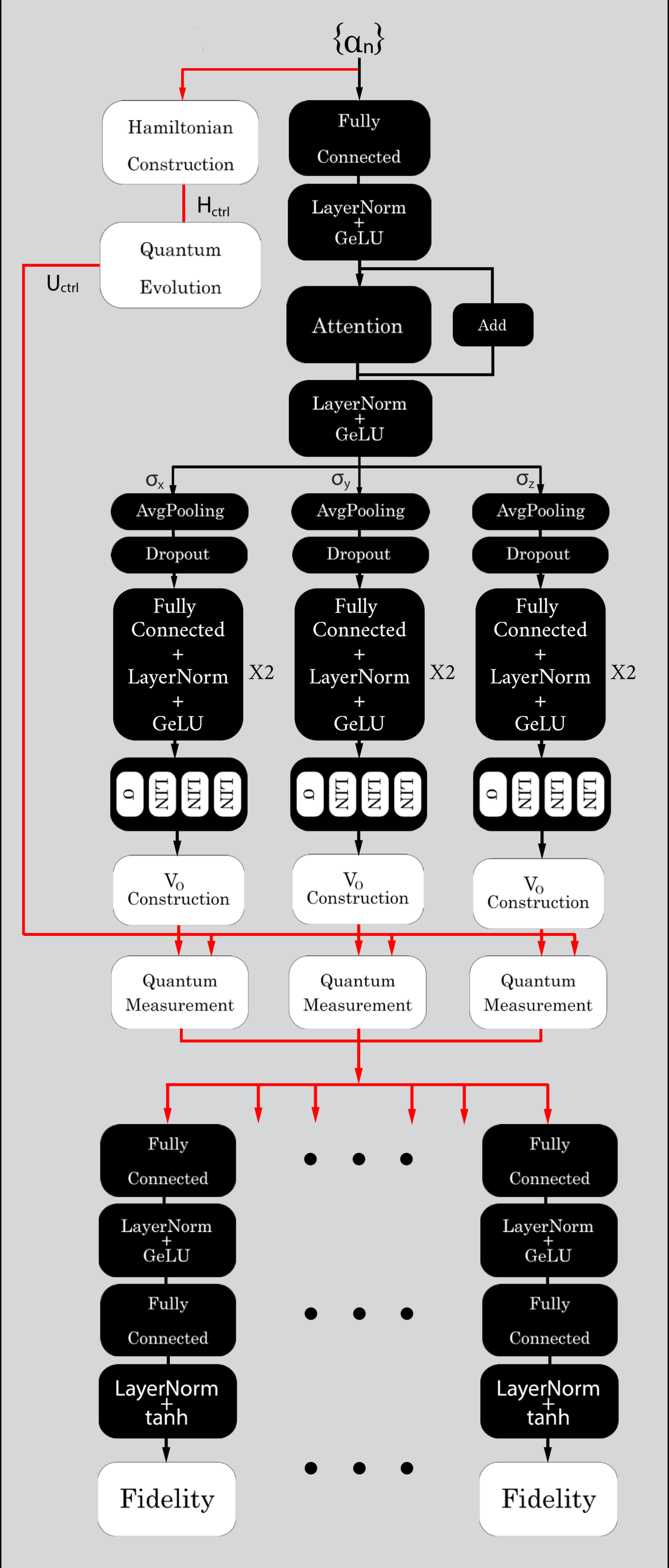
    }
    \caption{The transformer-based graybox architecture we used to model
    Markovian and non-Markovian open-system qubit dynamics.}
    \label{NNtransfin}
  \end{figure}

  A schematic overview of the ML architecture we use in this paper is shown in
  Fig.~\ref{NNtransfin}.

  \subsection{\label{sec:inputs}The network inputs}
  The ML model takes as input a set of parameters describing the Gaussian pulses
  applied along the $x$- and $y$-axes, see eqs.~\eqref{eq:cntrlham} and ~\eqref{eq:controls}.
  Each pulse $k$ (with $k=1,\dots,5$) is characterised by three parameters: its amplitude
  $A_{k}$, its relative position $\tau_{k}/T$, and its width $\sigma_{k}/ T$. We
  keep the positions $\tau_{k}$ and widths $\sigma_{k}$ fixed and identical
  across all input datapoints, while the amplitudes $A_{k}$ are randomized independently
  for each pulse. The total input thus consists of $10$ real numbers, which,
  together with the fixed $\tau_{k}$ and $\sigma_{k}$, completely characterize the
  drives. Prior to input, each parameter is normalized so that all entries lie in
  $[0,1]$, ensuring uniform scale across different datasets and preventing any
  single parameter from dominating the training.

  \subsection{The network outputs}
  The output of the model consists of six fidelity values, each corresponding to
  a target gate in the selected universal set of single-qubit operations, eq.~\eqref{eq:Qgates}.

  Internally, the model first computes a tomographically complete set of expectation
  values. Considering as initial states of the qubit each of the six eigenstates
  of the Pauli operators
  \begin{equation}
    \label{eq:initialstatesPopers}S = \left\{\,\ket{0_{x}},\ket{1_{x}},\; \ket{0_{y}}
    ,\ket{1_{y}},\; \ket{0_{z}},\ket{1_{z}}\,\right\},
  \end{equation}
  the expectation value $\avg{\sigma_{\alpha}(T)}_{\ket{\psi}}$ with
  $\alpha \in \{x, y, z\}$ is estimated, producing a set of $18$ real values.

  These expectation values are not the final output but are passed through a refinement
  section of the neural network. The refined representations are then used to
  calculate the six process matrix fidelities, one for each target gate in the selected
  universal set. These fidelities constitute the final output of the model.

  \subsection{The whitebox layers}
  The whitebox layers are fixed, physics-informed layers that explicitly implement
  the known parts of the qubit dynamics~\cite{youssri1}. These layers operate as
  follows:
  \paragraph{Hamiltonian construction.}
  The pulse parameters provided as input to the model are first used to compute
  a discretized control waveform. The total evolution time $T$ is divided into $M
  =2000$ equal time steps, and at each time $t_{k}$, the corresponding control amplitudes
  $\{f_{x}(t_{k}), f_{y}(t_{k})\}_{k=0}^{M-1}$ are evaluated from the input parameters.

  Based on these values, the instantaneous control Hamiltonian is constructed at
  each time step:
  \[
    H_{\rm ctrl}(t_{k}) = f_{x}(t_{k})\,\sigma_{x}\;+\; f_{y}(t_{k})\,\sigma_{y}.
  \]
  This step involves no trainable parameters and forms the basis for computing the
  controlled evolution.

  \paragraph{Controlled evolution.}
  Given the sequence $\{H_{\rm ctrl}(t_{k})\}$, the time-ordered exponential is approximated
  as
  \begin{equation}
    U_{\rm ctrl}~=~ \mathcal{T}_{+}\exp\!\Bigl[-i\!\int_{0}^{T}H_{\rm ctrl}(s)\,
    ds\Bigr] ~\approx~ \prod_{k=0}^{M-1}e^{-i\,H_{\rm ctrl}(t_k)\,\Delta T},
  \end{equation}
  where $\Delta T = T/M$. As $M$ increases, this product converges to the exact unitary
  generated by the control fields. The resulting unitary $U_{\rm ctrl}$ feeds
  into the Quantum Measurement whitebox layer.
  \paragraph{$V_{O}$ assembly.}
  The blackbox modules supply four real parameters $\{\mu,\theta,\psi,\Delta\}$
  that are used to calculate the noise-encoding operator $V_{O}$ , as expressed
  by Eq. \ref{VO}.
  \paragraph{Quantum Measurement.}
  For each input state $\rho(0)\in S$, eq.~\eqref{eq:initialstatesPopers}, the expectation
  value, eq.~\eqref{expect} is calculated. There is a Quantum Measurement layer for
  each Pauli operator, thus, in total, 18 expectation values are calculated, one
  for each combination of the six state preparations and three Pauli measurements.
  \paragraph{Fidelity.}
  In the final whitebox layers the model uses a representation of the 18 predicted
  expectation values to construct the process matrix. After that, the process matrix
  fidelity is computed with respect to each target gate, producing six real
  values.

  Embedding these whitebox steps directly into the architecture guarantees exact
  enforcement of unitary evolution and measurement evaluations, reducing the burden
  on the trainable components and ensuring physically consistent predictions under
  noise.

  \subsection{The blackbox layers}
  We use a lightweight Transformer-style encoder \cite{attention} as the core of
  the blackbox part of the neural network architecture. The blackbox layers
  process the input control pulses in three main stages:

  \begin{itemize}
    \item \textbf{Shared dense projection.} The normalized pulse parameters are
      first passed through a shared linear projection layer with 32 units. This projects
      the input into a latent representation that facilitates learning the underlying
      structure of the pulse sequence.

    \item \textbf{Transformer encoder.} The projected sequence is processed by a
      lightweight self-attention encoder consisting of a multi-head attention
      block (with two attention heads) followed by a feed-forward network. This encoder
      extracts contextual relationships between all control pulses, refining the
      representation to capture correlations relevant to noise-induced effects.

    \item \textbf{Output branches.} The encoded sequence feeds three parallel
      output branches, one for each Pauli observable. In each branch, the global
      features are aggregated and passed through two dense layers. Each branch
      then outputs:
      \begin{itemize}
        \item A linear layer with three outputs defining the parameters $\psi$,
          $\theta$, and $\Delta$ for the unitary part of $V_{O}$.

        \item A sigmoid node producing the parameter $\mu$, constrained to
          $[0,1]$.
      \end{itemize}
  \end{itemize}

  The outputs from these branches parameterize the noise-averaged operation for each
  observable. The final expectation values are refined using additional heads
  that adjust the raw predictions and feed them into custom layers that compute
  the process fidelities for the gates eqs.~\eqref{eq:Qgates}. To improve fidelity
  predictions, the model includes a dedicated refinement head for each target
  gate. Each head independently adjusts the raw expectation values to account for
  gate-specific noise sensitivities and control features.

  A more detailed description of the model architecture is provided in Appendix \ref{sec:networkspecifics}.

  For training the neural network we utilize the sum of the mean squared errors (MSE)
  of the 6 fidelity outputs, as cost function. The model's parameters, are
  optimized to minimize this cost function. To this end, we employ the Adam
  optimizer~\cite{adam}. It is worth noting that the whitebox components of the model
  do not contain any trainable parameters. As a result, the training process enforces
  the outputs of the blackbox components to be compatible with the constraints imposed
  by the whitebox layers. This ensures that the overall model produces
  physically meaningful results. Once training is complete, the model should be
  capable of accurately predicting the outputs for the training examples. To verify
  that the model can generalize to unseen examples, after training, the model is
  tested on data that wasn't used during training. The MSE is calculated for
  this test subset, and the results are compared with the MSE on the training subset.
  These results are reported in tables~\ref{table:MSE_attn_RTN} and~\ref{table:MSE_attn_OU}.

  The numerical implementation of the graybox model and of the quantum evolution
  of the system was developed in Python~\cite{python}, using the TensorFlow framework
  along with its high-level Keras API~\cite{tensorflow, keras}.

  \subsection{\label{sec:datagen}Data Generation}
  In order to generate the necessary data, we solve Eq.~\eqref{standard_exp}
  numerically, using a Monte Carlo method with total evolution time $T = 3.2\,\mu
  s$, a discretized time interval comprising $M = 3 000$ time steps, and $K = 200
  0$ independent noise realizations for statistical averaging. This setup allows
  for a detailed and accurate representation of the behavior of the qubit under
  noisy conditions.

  Our case studies are designed to explore the system's response across a
  variety of noise coupling regimes. In all scenarios, the environmental noise
  features a fixed rate $\gamma = k/2 = 1\,\text{MHz}$, ensuring consistency in the
  noise dynamics while varying the noise coupling strength.

  For both the RTN and OU cases, we generated a dataset for each of the
  following coupling values: $g = 0.1$ MHz, $g = 0.2$ MHz, $g = 0.3$ MHz,
  $g = 0.4$ MHz, and $g = 0.5$ MHz. For the first four coupling strengths, each dataset
  contains $5000$ data points, while for $g = 0.5$ MHz we generated a larger
  dataset of $10000$ data points. Each dataset was divided into training and testing
  sets following an 80:20 ratio.

  \begin{table*}
    [!ht]
    \centering
    \footnotesize
    \begin{tabular}{|c|c||c|c|c|c|c|c|}
      \hline
      $g$                      &            & I                    & X                    & Y                    & Z                    & H                    & $R_{X}\!\bigl(\frac{\pi}{4}\bigr)$ \\
      \hline
      \hline
      \multirow{2}{*}{0.1 MHz} & MSE$_{tr}$ & $1.34\times 10^{-5}$ & $1.57\times 10^{-5}$ & $1.53\times 10^{-5}$ & $1.58\times 10^{-5}$ & $1.39\times 10^{-5}$ & $1.38\times 10^{-5}$               \\
                               & MSE$_{te}$ & $8.16\times 10^{-6}$ & $1.28\times 10^{-5}$ & $8.83\times 10^{-6}$ & $1.51\times 10^{-5}$ & $9.07\times 10^{-6}$ & $1.06\times 10^{-5}$               \\
      \hline
      \multirow{2}{*}{0.2 MHz} & MSE$_{tr}$ & $7.66\times 10^{-5}$ & $9.65\times 10^{-5}$ & $9.26\times 10^{-5}$ & $6.82\times 10^{-5}$ & $8.89\times 10^{-5}$ & $6.18\times 10^{-5}$               \\
                               & MSE$_{te}$ & $7.16\times 10^{-5}$ & $5.59\times 10^{-5}$ & $6.19\times 10^{-5}$ & $6.36\times 10^{-5}$ & $7.81\times 10^{-5}$ & $5.08\times 10^{-5}$               \\
      \hline
      \multirow{2}{*}{0.3 MHz} & MSE$_{tr}$ & $1.26\times 10^{-4}$ & $1.60\times 10^{-4}$ & $1.54\times 10^{-4}$ & $1.40\times 10^{-4}$ & $1.54\times 10^{-4}$ & $1.29\times 10^{-4}$               \\
                               & MSE$_{te}$ & $1.25\times 10^{-4}$ & $1.38\times 10^{-4}$ & $1.40\times 10^{-4}$ & $1.33\times 10^{-4}$ & $1.38\times 10^{-4}$ & $1.33\times 10^{-4}$               \\
      \hline
      \multirow{2}{*}{0.4 MHz} & MSE$_{tr}$ & $2.12\times 10^{-4}$ & $2.48\times 10^{-4}$ & $2.55\times 10^{-4}$ & $2.14\times 10^{-4}$ & $2.26\times 10^{-4}$ & $2.16\times 10^{-4}$               \\
                               & MSE$_{te}$ & $2.12\times 10^{-4}$ & $3.05\times 10^{-4}$ & $2.85\times 10^{-4}$ & $2.66\times 10^{-4}$ & $2.81\times 10^{-4}$ & $2.26\times 10^{-4}$               \\
      \hline
      \multirow{2}{*}{0.5 MHz} & MSE$_{tr}$ & $4.58\times 10^{-4}$ & $5.24\times 10^{-4}$ & $5.37\times 10^{-4}$ & $4.61\times 10^{-4}$ & $4.99\times 10^{-4}$ & $4.68\times 10^{-4}$               \\
                               & MSE$_{te}$ & $4.76\times 10^{-4}$ & $5.90\times 10^{-4}$ & $5.59\times 10^{-4}$ & $5.58\times 10^{-4}$ & $5.72\times 10^{-4}$ & $4.89\times 10^{-4}$               \\
      \hline
    \end{tabular}
    \caption{\label{table:MSE_attn_RTN}Training (MSE$_{tr}$) and testing (MSE$_{te}$)
    mean squared error values for each target gate at different coupling strengths.}
  \end{table*}

  \section{\label{sec:results}Results}
  \subsection{\label{RTNcase}RTN case}
  \subsubsection{\label{sec:RTNtraining}Training}
  We trained separate instances of the graybox model on each of the datasets
  generated for the various values of $g$. The final train and test loss quantifying
  the performance of the prediction of each gate fidelity is summarized in Table~\ref{table:MSE_attn_RTN}.

  It shows that the graybox attention-based model maintains low training and testing
  MSE values across all gates and coupling strengths. Furthermore, the good alignment
  of training and testing MSE indicates stable training and no apparent overfitting.
  To better interpret these results, we also report in Table~\ref{table:prediction_errors_attn}
  the corresponding average prediction error for each coupling strength,
  calculated as the square root of the average MSE across all gates. This offers
  an intuitive measure of how accurately the model generalizes to unseen examples.
  It is apparent that the prediction error increases with increasing coupling
  strength. This trend shows that the model, while remaining reliable with an
  error of the order of $10^{-2}-10^{-3}$, faces growing difficulty in capturing
  more complex dynamics at stronger coupling.
  \begin{table}[H]
    \centering
    \footnotesize
    \begin{tabular}{|c|c|}
      \hline
      $g$ (MHz) & Prediction Error \\
      \hline
      \hline
      0.1       & 0.0033           \\
      \hline
      0.2       & 0.0080           \\
      \hline
      0.3       & 0.0116           \\
      \hline
      0.4       & 0.0162           \\
      \hline
      0.5       & 0.0233           \\
      \hline
    \end{tabular}
    \caption{\label{table:prediction_errors_attn}Average prediction error corresponding
    to the square root of the mean test MSE over the gates, for each coupling
    strength $g$.}
  \end{table}

  \subsubsection{\label{sec:RTNoptimalcontrol}Optimal Control}
  After training, the graybox model was used as a quantum system emulator in the
  optimal control protocol described in Subsection~\ref{sec:optimalcontrol}.

\begin{figure}[!ht]
    \centering
    \includegraphics[]{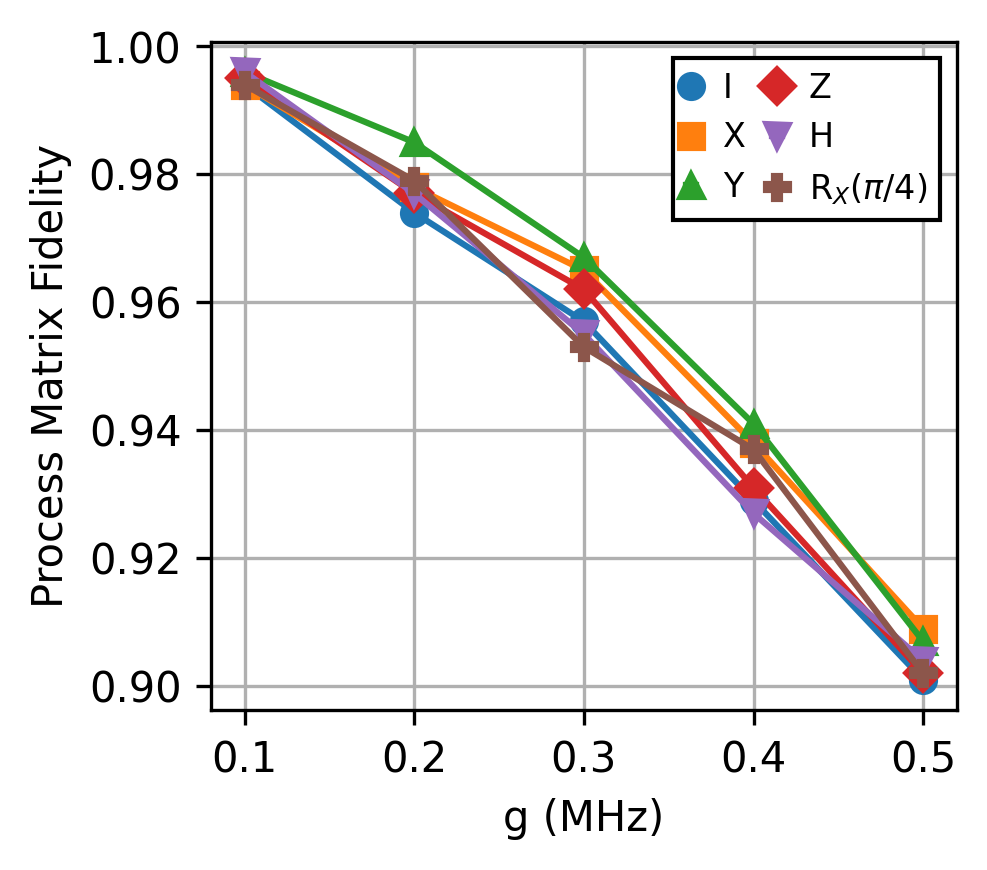}
    \caption{\label{fig:fid_vs_g_RTN}Process matrix fidelity versus RTN noise coupling
    strength $g$ for each gate in the universal set $S$.}
  \end{figure}

  The resulting optimized fidelities are reported in Fig.~\ref{fig:fid_vs_g_RTN}
  and Tab.~\ref{table: fidelities_attn_RTN} for the various values of $g$. The fidelity
  values reported here are obtained by feeding the pulse sequences generated by the
  optimal control protocol into a Monte Carlo simulation of the system.
  \begin{table}[!ht]
    \centering
    \footnotesize
    \begin{tabular}{|c||c|c|c|c|c|}
      \hline
                                         & $g = 0.1$ & $g = 0.2$ & $g = 0.3$ & $g = 0.4$ & $g = 0.5$ \\
      \hline
      \hline
      I                                  & 0.994     & 0.974     & 0.957     & 0.929     & 0.901     \\
      \hline
      X                                  & 0.994     & 0.978     & 0.965     & 0.938     & 0.909     \\
      \hline
      Y                                  & 0.996     & 0.985     & 0.967     & 0.941     & 0.907     \\
      \hline
      Z                                  & 0.995     & 0.977     & 0.962     & 0.931     & 0.902     \\
      \hline
      H                                  & 0.996     & 0.977     & 0.955     & 0.927     & 0.904     \\
      \hline
      $R_{X}\!\bigl(\frac{\pi}{4}\bigr)$ & 0.994     & 0.979     & 0.953     & 0.937     & 0.902     \\
      \hline
    \end{tabular}
    \caption{Process matrix fidelity values for each universal gate at different
    coupling strengths $g$.}
    \label{table: fidelities_attn_RTN}
  \end{table}
  These results show that the graybox model maintains high process fidelities
  across a broad range of noise coupling strengths $g$. As expected, the
  fidelities decrease as $g$ increases, reflecting the stronger impact of noise on
  the system's dynamics. For $g = 0.1$, the model achieves fidelities above 99\%
  for all gates, indicating that the model accurately captures the system's behavior
  in low-noise conditions and the optimization finds control sequences which are
  highly effective. As the coupling strength increases, fidelities gradually decline,
  but remain above 90\% even for $g = 0.5$. The results also highlight small
  variations in fidelity across different gates at the same coupling strength. Overall,
  the attention-based graybox model demonstrates robust performance, providing effective
  control solutions tailored to different gates even under challenging noisy conditions.

  \begin{table*}
    [!ht]
    \centering
    \footnotesize
    \begin{tabular}{|c|c||c|c|c|c|c|c|}
      \hline
      $g$                      &            & I                    & X                    & Y                    & Z                    & H                    & $R_{X}\!\bigl(\frac{\pi}{4}\bigr)$ \\
      \hline
      \hline
      \multirow{2}{*}{0.1 MHz} & MSE$_{tr}$ & $2.69\times 10^{-6}$ & $3.02\times 10^{-6}$ & $2.95\times 10^{-6}$ & $2.66\times 10^{-6}$ & $2.73\times 10^{-6}$ & $2.69\times 10^{-6}$               \\
                               & MSE$_{te}$ & $3.29\times 10^{-6}$ & $3.47\times 10^{-6}$ & $3.48\times 10^{-6}$ & $3.11\times 10^{-6}$ & $3.06\times 10^{-6}$ & $3.28\times 10^{-6}$               \\
      \hline
      \multirow{2}{*}{0.2 MHz} & MSE$_{tr}$ & $6.83\times 10^{-5}$ & $6.66\times 10^{-5}$ & $6.83\times 10^{-5}$ & $8.04\times 10^{-5}$ & $7.41\times 10^{-5}$ & $6.74\times 10^{-5}$               \\
                               & MSE$_{te}$ & $4.98\times 10^{-5}$ & $4.98\times 10^{-5}$ & $6.82\times 10^{-5}$ & $6.32\times 10^{-5}$ & $4.49\times 10^{-5}$ & $5.90\times 10^{-5}$               \\
      \hline
      \multirow{2}{*}{0.3 MHz} & MSE$_{tr}$ & $1.31\times 10^{-4}$ & $1.74\times 10^{-4}$ & $1.49\times 10^{-4}$ & $1.40\times 10^{-4}$ & $1.42\times 10^{-4}$ & $1.45\times 10^{-4}$               \\
                               & MSE$_{te}$ & $1.15\times 10^{-4}$ & $1.35\times 10^{-4}$ & $1.29\times 10^{-4}$ & $1.11\times 10^{-4}$ & $1.33\times 10^{-4}$ & $1.27\times 10^{-4}$               \\
      \hline
      \multirow{2}{*}{0.4 MHz} & MSE$_{tr}$ & $2.57\times 10^{-4}$ & $3.16\times 10^{-4}$ & $2.95\times 10^{-4}$ & $2.79\times 10^{-4}$ & $2.84\times 10^{-4}$ & $2.72\times 10^{-4}$               \\
                               & MSE$_{te}$ & $2.38\times 10^{-4}$ & $3.26\times 10^{-4}$ & $2.92\times 10^{-4}$ & $2.71\times 10^{-4}$ & $2.74\times 10^{-4}$ & $2.90\times 10^{-4}$               \\
      \hline
      \multirow{2}{*}{0.5 MHz} & MSE$_{tr}$ & $3.75\times 10^{-4}$ & $4.72\times 10^{-4}$ & $4.67\times 10^{-4}$ & $4.08\times 10^{-4}$ & $4.30\times 10^{-4}$ & $3.85\times 10^{-4}$               \\
                               & MSE$_{te}$ & $4.04\times 10^{-4}$ & $5.04\times 10^{-4}$ & $5.11\times 10^{-4}$ & $4.61\times 10^{-4}$ & $4.54\times 10^{-4}$ & $4.12\times 10^{-4}$               \\
      \hline
    \end{tabular}
    \caption{Training (MSE$_{tr}$) and testing (MSE$_{te}$) mean squared error
    values for each target gate at different coupling strengths $g$ in the
    Ornstein-Uhlenbeck case.}
    \label{table:MSE_attn_OU}
  \end{table*}
  \subsection{\label{sec:OUcase}OU case}
  \subsubsection{\label{sec:OUtraining}Training}
  Following the same strategy used for the RTN noise we train a different instance
  of the graybox model on each of the datasets generated for the OU case for different
  couplings $g$.

  The prediction performance for each gate and coupling value is summarized in Table~\ref{table:MSE_attn_OU},
  which reports the final training and testing MSE values for each gate. Results
  show that the graybox attention-based model maintains consistently low training
  and testing MSE values across all gates and coupling strengths also for the Ornstein-Uhlenbeck
  noise case. As before, we also report in Table~\ref{table:prediction_errors_attn_OU}
  the corresponding average prediction error for each coupling strength. A
  comparison with the RTN results shows that training the OU model achieves very
  similar prediction accuracy across all coupling strengths, demonstrating the robustness
  of the graybox attention-based approach accross different noise types.
  \begin{table}[H]
    \centering
    \footnotesize
    \begin{tabular}{|c|c|}
      \hline
      $g$ (MHz) & Prediction Error \\
      \hline
      \hline
      0.1       & 0.0018           \\
      \hline
      0.2       & 0.0075           \\
      \hline
      0.3       & 0.0112           \\
      \hline
      0.4       & 0.0168           \\
      \hline
      0.5       & 0.0214           \\
      \hline
    \end{tabular}
    \caption{Average prediction error corresponding to the square root of the
    mean test MSE over all the gates in $S$, for each coupling strength $g$ in the
    Ornstein-Uhlenbeck case.}
    \label{table:prediction_errors_attn_OU}
  \end{table}

  \subsubsection{\label{sec:OUoptimalcontrol}Optimal Control}

  After training, the attention-based graybox model is used as an emulator for the
  optimal control stage.

\begin{figure}[!ht]
    \centering
    \includegraphics[]{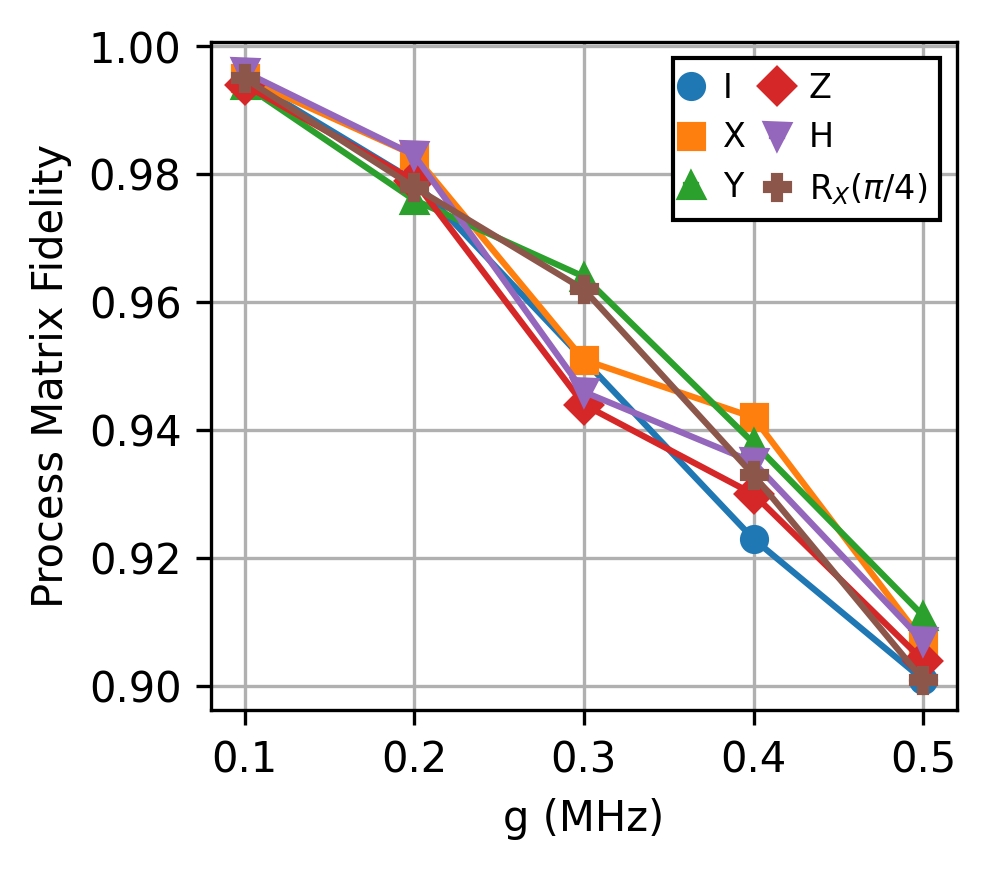}
    \caption{\label{fig:fid_vs_g_OU}Process matrix fidelity versus OU noise coupling
    strength $g$ for each gate in the universal set $S$.}
  \end{figure}

  Figure~\ref{fig:fid_vs_g_OU} and Tab.~\ref{table:fidelities_attn_OU} report
  the optimized process matrix fidelities obtained as result of the optimal
  control procedure.
  \begin{table}[H]
    \centering
    \footnotesize
    \begin{tabular}{|c||c|c|c|c|c|}
      \hline
                                                    & $\textbf{g}= 0.1$ & $\textbf{g}= 0.2$ & $\textbf{g}= 0.3$ & $\textbf{g}= 0.4$ & $\textbf{g}= 0.5$ \\
      \hline
      \hline
      $\textbf{I}$                                  & 0.995             & 0.979             & 0.951             & 0.923             & 0.901             \\
      \hline
      $\textbf{X}$                                  & 0.995             & 0.983             & 0.951             & 0.942             & 0.907             \\
      \hline
      $\textbf{Y}$                                  & 0.994             & 0.976             & 0.964             & 0.938             & 0.911             \\
      \hline
      $\textbf{Z}$                                  & 0.994             & 0.979             & 0.944             & 0.930             & 0.904             \\
      \hline
      $\textbf{H}$                                  & 0.996             & 0.983             & 0.946             & 0.935             & 0.907             \\
      \hline
      $\boldsymbol{R_X\!\bigl(\frac{\pi}{4}\bigr)}$ & 0.995             & 0.978             & 0.962             & 0.933             & 0.901             \\
      \hline
    \end{tabular}
    \caption{Process matrix fidelity values for each target gate at different
    coupling strengths $g$ in the OU case.}
    \label{table:fidelities_attn_OU}
  \end{table}
  The results are comparable to the one obtained in the RTN noise case. For
  $g = 0.1$, all gates reach fidelities above 99\%. As the coupling strength increases,
  the fidelities gradually decrease, reaching values around 90\% for $g = 0.5$. Notably,
  the performance remains consistent across different gates, with minor variations
  that reflect specific gate sensitivities to noise and control imperfections.
  Despite the differences between noise models, the attention-based graybox approach
  remains robust in both settings. The optimization procedure, using the ML model
  as emulator, continues to generate effective pulse sequences that achieve high
  process fidelities in most cases. Nonetheless, at very high coupling values
  where decoherence is high, neither noise model offers a straightforward path to
  sustaining fidelity near $1.0$ and both may require larger training datasets,
  longer training times, or more sophisticated control strategies.

  \section{\label{sec:conclusions}Conclusions}
  In this paper, we presented an implementation of a graybox machine-learning
  framework based on transformers for modeling and controlling open quantum
  systems. We focused our study on a single qubit whose dynamics can be non-Markovian.
  Our results show that the model accurately predicts the qubit's dynamics
  across a wide range of coupling strengths. In addition to RTN, we analyzed the
  a system subject to Ornstein-Uhlenbeck (OU) noise, showing that the non-Gaussian
  character of RTN does not appreciably alter the optimization results compared to
  its Gaussian OU counterpart.

  A key outcome of this work is the demonstration that the trained attention-based
  graybox model can be repurposed as a fast emulator for optimal control,
  enabling the design of pulse trains that implement a universal set of quantum
  gates with high fidelity even under increasing noise coupling. Our results show
  that the method retains good predictive accuracy and control performance where
  purely analytical or fully whitebox methods may fail or require additional approximations.

  A significant aspect of this work is the integration of an attention-based neural
  network architecture within the graybox framework. This design enables stable
  and efficient training, with fast convergence times and effective GPU parallelization
  during both training and inference. The architecture includes dedicated
  fidelity prediction heads, each specializing in refining the estimation of the
  process fidelity of a specific target gate. This structure allows the model to
  focus on gate-specific control features and noise sensitivities, leading to high
  process fidelities in optimal control tasks.

  Several promising directions for future research remain:
  \begin{enumerate}
    \item \textbf{Scaling with noise strength and dataset size.} Stronger noise
      regimes demand larger datasets to capture complex system-environment
      correlations accurately. Future work can explore systematic strategies for
      dataset generation and optimal parameter sampling, ensuring that the model
      size and training costs remain manageable even in with higher values of
      the coupling.

    \item \textbf{Adaptation to real-device constraints.} Practical quantum
      hardware may impose constraints on accessible control pulses. The graybox framework
      can be adapted to these limitations by embedding hardware-specific constraints
      during training, enabling device-tailored optimal control solutions.

    \item \textbf{Generalization to multiqubit and multilevel systems.} Although
      this work focused on a single qubit, the method naturally extends to multiqubit
      registers and multilevel systems. The software developed here can be
      adapted for two-qubit gates constructed from two-level unitary blocks~\cite{falci3}.
      Future work should explore scaling properties and how additional whitebox layers
      for multi-qubit Hamiltonians could improve training efficiency.
  \end{enumerate}

  In summary, this paper offers a demonstration that the graybox machine-learning
  paradigm offers a powerful and flexible framework for modeling open quantum
  dynamics and designing optimal control protocols in noisy environments. Beyond
  gate design, the same approach holds promise for noise spectroscopy, quantum error
  mitigation, and other quantum control challenges that demand adaptive, data-driven
  solutions. As quantum hardware scales and environmental noise becomes
  increasingly complex, we anticipate that hybrid graybox strategies,
  strengthened by modern neural architectures such as attention mechanisms, will
  become useful tools for pushing the frontiers of quantum technologies.

  \begin{acknowledgments}
    RC and SM acknowlwdge support from the ICSC - Centro Nazionale di Ricerca in
    High-Performance Computing, Big Data and Quantum Computing. EP and LG acknowledge
    support by the PNRR MUR project PE0000023-NQSTI. GAF is supported from the
    PRIN 2022WKCJRT project SuperNISQ and from the University of Catania, Piano Incentivi
    Ricerca di Ateneo 2024-26, project QTCM. EP acknowledges the COST Action SUPERQUMAP
    (CA 21144).
  \end{acknowledgments}


  \appendix
  \section{\label{app:hamderivation}Hamiltonian derivation}
  Here we show how to derive the Hamiltonian in Eq.~\eqref{totalham}~\cite{youssri1}\cite{paz-silva}.
  We begin with a general formulation of the Hamiltonian for a single qubit interacting
  with an arbitrary environment (or "bath") under the influence of external control
  and subsequently introduce some simplifying assumptions \cite{paz-silva}. In
  the laboratory frame, the dynamics of a composite quantum system, consisting
  of a qubit (system) interacting with its environment, are governed by the following
  Hamiltonian:
  \begin{equation*}
    \Htot(t) = \Hs+ \Hb+ \Hsb(t) + \Hctrl^{\text{lab}}( t ),
  \end{equation*}
  where $\Hs$ and $\Hb$ describe the intrinsic free dynamics of the system and
  bath, respectively. The interaction between the system and its environment is
  captured by $\Hsb(t)$, while the control Hamiltonian $\Hctrl^{\text{lab}}(t)$,
  acting exclusively on the system's Hilbert space $\mathcal{H}_{S}$, represents
  externally applied control operations.

  In natural units ($\hbar = 1$), the internal system Hamiltonian and the system-bath
  interaction are expressed as:
  \begin{equation*}
    \Hs= \frac{\Omega}{2}\sigma_{z}, \quad \Hsb(t) = \sum_{\alpha = x,y,z}\sigma_{\alpha}
    \otimes B_{\alpha}^{0}(t),
  \end{equation*}

  where $\Omega$ represents the energy splitting of the qubit, and $B_{\alpha}^{0}
  (t) = B_{\alpha}^{0\dagger}(t)$ are time-dependent bath operators characterizing
  the environmental degrees of freedom.

  A notable aspect of this framework lies in the fact that bath operators have both
  quantum and classical components. This is formally described as:
  \begin{equation*}
    B_{\alpha}^{0}(t) = \tilde{B}_{\alpha}^{0}(t) + \beta_{\alpha}(t) I_{B},
  \end{equation*}
  where $\tilde{B}_{\alpha}^{0}(t)$ denotes the quantum (non-commuting) contribution,
  $\beta_{\alpha}(t)$ is a classical stochastic process, and $I_{B}$ is the identity
  operator on the bath's Hilbert space $\mathcal{H}_{B}$. In the purely
  classical limit (which is the one we explore in this paper), the quantum component
  $\tilde{B}_{\alpha}^{0}(t)$ vanishes, and the bath operators reduce to
  $B_{\alpha}^{0}(t) = \beta_{\alpha}(t) I_{B}$.

  We specifically focus on dephasing noise, which arises from interactions coupling
  exclusively to the axis along which the qubit's energy eigenstates are defined.
  As is most often the case, we take said axis to be $\sigma_{z}$. In the
  interaction picture, where the dynamics are described relative to the internal
  Hamiltonian of the system and bath, $\Hs+ \Hb$, the evolution of the qubit-bath
  system is governed by the interaction Hamiltonian:
  \begin{equation*}
    \Hi(t) = \sum_{\alpha = x, y, z}\sigma_{\alpha}(t) \otimes B_{\alpha}(t) + \Hctrl
    (t).
  \end{equation*}
  The relevant terms are explicitly given by:
  \begin{align*}
    B_{\alpha}(t)      & \equiv e^{i H_B t}B_{\alpha 0}(t) e^{-i H_B t},                                                    \\
    \Hctrl(t)          & \equiv e^{i \frac{\Omega t}{2} \sigma_z}\Hctrl^{\text{lab}}(t) e^{-i \frac{\Omega t}{2} \sigma_z}, \\
    \sigma_{\alpha}(t) & \equiv e^{i \frac{\Omega t}{2} \sigma_z}\sigma_{\alpha}e^{-i \frac{\Omega t}{2} \sigma_z}.
  \end{align*}

  In this formalism, the bath operators $B_{\alpha}(t)$ evolve due to the bath's
  free dynamics governed by $\Hb$, while the control Hamiltonian $\Hctrl(t)$
  accounts for the modulation of the laboratory-frame control fields by the
  qubit's internal dynamics, characterized by the energy splitting $\Omega$.
  Similarly, the qubit operators $\sigma_{\alpha}(t)$ transform under the influence
  of the system Hamiltonian $\Hs= \frac{\Omega}{2}\sigma_{z}$, introducing time-dependent
  behavior. In the specific case where the qubit is in the \textit{driftless
  regime}, characterized by $\Omega = 0$, the qubit's intrinsic energy splitting
  vanishes. As a result, the interaction-picture Hamiltonian simplifies
  significantly, reducing to:
  \begin{equation*}
    \Hi^{(\Omega = 0)}(t) = \sum_{\alpha = x, y, z}\sigma_{\alpha}\otimes B_{\alpha}
    (t) + \Hctrl(t).
  \end{equation*}
  Here, the absence of the qubit's free evolution ($\Hs= 0$) implies that the Pauli
  operators $\sigma_{\alpha}$ no longer acquire any time dependence.
  Consequently, the interaction Hamiltonian retains its static form in the qubit
  subspace, while the bath operators $B_{\alpha}(t)$ still evolve under the
  influence of the bath's internal dynamics governed by $\Hb$. In our analysis,
  we use $\Omega=0$ and a classical stochastic noise process directed along the
  $z$ axis. Thus, the full Hamiltonian of our case-study is reduced to:
  \begin{equation*}
    H(t) \equiv \Htot(t) = \Hctrl(t) + H_{1}(t), \label{eq:total_Hamiltonian}
  \end{equation*}
  where the control Hamiltonian $\Hctrl(t)$ is defined by:

  \begin{equation*}
    \Hctrl(t) = f_{x}(t)\sigma_{x}+ f_{y}(t)\sigma_{y}, \label{eq:control_Hamiltonian}
  \end{equation*}
  and the interaction with the noise is represented by:

  \begin{equation*}
    H_{1}(t) = g \beta(t) \sigma_{z}. \label{eq:interaction_Hamiltonian}
  \end{equation*}
  In these expressions, $f_{x}(t)$ and $f_{y}(t)$ are time-dependent control
  fields applied along the $x$ and $y$ axes of the qubit, respectively; $g$
  denotes the coupling strength between the qubit and the stochastic noise
  process $\beta(t)$; and $\beta(t)$ is either modeled as a Random Telegraph Noise
  (RTN) or an Ornstein-Uhlenbeck (OU) process.

  \section{\label{sec:networkspecifics}Neural network specifics}

  In this section, we briefly present the main components and features of the blackbox
  part of the hybrid architecture proposed in this work.

  \paragraph{Scaled dot-product attention.}
  At the heart of the transformer-based blackbox is the \emph{attention} mechanism.
  In this approach, the network learns how much attention each part of the control
  pulse sequence should pay to every other part. By computing these attention
  scores, the model can combine information from all time steps, capturing both
  short and long-range correlations in the pulse train. This direct, context-wide
  perspective overcomes the vanishing-gradient and limited receptive-field
  issues afflicting a wide variety of architecture types, enabling stable training
  and robust modeling even for long control sequences~\cite{attention}.

  \paragraph{Multi-head attention.}
  To increase flexibility, the transformer uses \emph{multi-head attention}, which
  splits the calculation into several parallel heads. Each attention head can
  focus on different features in the sequence - for example, local overlaps or global
  trends - and these diverse views are then combined into a single, rich
  representation. This makes the model more effective at learning how the entire
  control signal influences the qubit's dynamics.

  \paragraph{Parallelism and efficiency.}
  Self‐attention conducts all dot‐product similarity computations in parallel-both
  during training and inference, exploiting modern GPU tensor cores to their full
  extent. This parallel structure dramatically accelerates both forward passes
  and gradient calculations, making attention‐based encoders highly scalable for
  large batch sizes and long control sequences~\cite{effattention}.

  The Transformer-based architecture incorporates modern design features that ensure
  stable and fast training complete (a single run takes less than an hour) and achieve
  low final MSE. These results are enabled by the components showcased below.

  \paragraph{Residual connections.}
  Residual skip connections (\texttt{Add} layers) between the self-attention and
  feed-forward blocks allow gradients to bypass intermediate transformations~\cite{he2016deep}.
  This stabilizes training by mitigating vanishing or exploding gradients,
  contributing directly to faster and more robust convergence.

  \paragraph{Layer normalization.}
  Layer normalization (\texttt{LayerNorm}) is applied after attention and feed-forward
  sublayers. By re-centering and rescaling intermediate representations, it
  prevents internal covariate shifts, accelerating learning dynamics for long pulse
  sequences~\cite{ba2016layer}.

  \paragraph{GeLU activations.}
  The use of Gaussian Error Linear Units (GeLU) instead of ReLU provides
  smoother gating of neuron outputs, which empirically improves the optimizer's ability
  to fit fine-grained correlations in the qubit's control pulses and noise interactions,
  helping to lower final MSE~\cite{hendrycks2016gaussian}.

  \paragraph{Global average pooling.}
  After self-attention, a \texttt{GlobalAveragePooling1D} operation compresses each
  sequence's context into fixed-size feature vectors. This pooling allows the
  model to adapt seamlessly to pulse trains of varying lengths, increasing practical
  flexibility and numerical stability~\cite{lin2014network}.

  \paragraph{Dropout regularization.}
  Dropout is applied after pooling to randomize active units during training, reducing
  overfitting and encouraging the network to learn more generalizable features~\cite{srivastava2014dropout}.

  \paragraph{Warmup and cosine decay scheduling.}
  The training uses a warmup phase in which the learning rate linearly increases
  to a peak, followed by a cosine decay. This avoids large, unstable parameter updates
  early on and enables smoother and faster convergence to a good local minimum~\cite{loshchilov2017sgdr}.\\

  Taken together, these enhancements result in a Transformer-based blackbox module
  with high expressive capacity and stable training dynamics.

  \bibliography{refs,S}

\providecommand{\noopsort}[1]{}\providecommand{\singleletter}[1]{#1}%
\begin{thebibliography}{54}%
\makeatletter
\providecommand \@ifxundefined [1]{%
 \@ifx{#1\undefined}
}%
\providecommand \@ifnum [1]{%
 \ifnum #1\expandafter \@firstoftwo
 \else \expandafter \@secondoftwo
 \fi
}%
\providecommand \@ifx [1]{%
 \ifx #1\expandafter \@firstoftwo
 \else \expandafter \@secondoftwo
 \fi
}%
\providecommand \natexlab [1]{#1}%
\providecommand \enquote  [1]{``#1''}%
\providecommand \bibnamefont  [1]{#1}%
\providecommand \bibfnamefont [1]{#1}%
\providecommand \citenamefont [1]{#1}%
\providecommand \href@noop [0]{\@secondoftwo}%
\providecommand \href [0]{\begingroup \@sanitize@url \@href}%
\providecommand \@href[1]{\@@startlink{#1}\@@href}%
\providecommand \@@href[1]{\endgroup#1\@@endlink}%
\providecommand \@sanitize@url [0]{\catcode `\\12\catcode `\$12\catcode
  `\&12\catcode `\#12\catcode `\^12\catcode `\_12\catcode `\%12\relax}%
\providecommand \@@startlink[1]{}%
\providecommand \@@endlink[0]{}%
\providecommand \url  [0]{\begingroup\@sanitize@url \@url }%
\providecommand \@url [1]{\endgroup\@href {#1}{\urlprefix }}%
\providecommand \urlprefix  [0]{URL }%
\providecommand \Eprint [0]{\href }%
\providecommand \doibase [0]{https://doi.org/}%
\providecommand \selectlanguage [0]{\@gobble}%
\providecommand \bibinfo  [0]{\@secondoftwo}%
\providecommand \bibfield  [0]{\@secondoftwo}%
\providecommand \translation [1]{[#1]}%
\providecommand \BibitemOpen [0]{}%
\providecommand \bibitemStop [0]{}%
\providecommand \bibitemNoStop [0]{.\EOS\space}%
\providecommand \EOS [0]{\spacefactor3000\relax}%
\providecommand \BibitemShut  [1]{\csname bibitem#1\endcsname}%
\let\auto@bib@innerbib\@empty
\bibitem [{\citenamefont {Ball}\ \emph {et~al.}(2021)\citenamefont {Ball},
  \citenamefont {Biercuk}, \citenamefont {Carvalho}, \citenamefont {Chen},
  \citenamefont {Hush}, \citenamefont {De~Castro}, \citenamefont {Li},
  \citenamefont {Liebermann}, \citenamefont {Slatyer}, \citenamefont {Edmunds}
  \emph {et~al.}}]{ball2021software}%
  \BibitemOpen
  \bibfield  {author} {\bibinfo {author} {\bibfnamefont {H.}~\bibnamefont
  {Ball}}, \bibinfo {author} {\bibfnamefont {M.~J.}\ \bibnamefont {Biercuk}},
  \bibinfo {author} {\bibfnamefont {A.~R.}\ \bibnamefont {Carvalho}}, \bibinfo
  {author} {\bibfnamefont {J.}~\bibnamefont {Chen}}, \bibinfo {author}
  {\bibfnamefont {M.}~\bibnamefont {Hush}}, \bibinfo {author} {\bibfnamefont
  {L.~A.}\ \bibnamefont {De~Castro}}, \bibinfo {author} {\bibfnamefont
  {L.}~\bibnamefont {Li}}, \bibinfo {author} {\bibfnamefont {P.~J.}\
  \bibnamefont {Liebermann}}, \bibinfo {author} {\bibfnamefont {H.~J.}\
  \bibnamefont {Slatyer}}, \bibinfo {author} {\bibfnamefont {C.}~\bibnamefont
  {Edmunds}}, \emph {et~al.},\ }\bibfield  {title} {\bibinfo {title} {Software
  tools for quantum control: Improving quantum computer performance through
  noise and error suppression},\ }\href@noop {} {\bibfield  {journal} {\bibinfo
   {journal} {Quantum Science and Technology}\ }\textbf {\bibinfo {volume}
  {6}},\ \bibinfo {pages} {044011} (\bibinfo {year} {2021})}\BibitemShut
  {NoStop}%
\bibitem [{\citenamefont {Abbott}\ \emph {et~al.}(2020)\citenamefont {Abbott},
  \citenamefont {Wechs}, \citenamefont {Horsman}, \citenamefont {Mhalla},\ and\
  \citenamefont {Branciard}}]{abbott2020communication}%
  \BibitemOpen
  \bibfield  {author} {\bibinfo {author} {\bibfnamefont {A.~A.}\ \bibnamefont
  {Abbott}}, \bibinfo {author} {\bibfnamefont {J.}~\bibnamefont {Wechs}},
  \bibinfo {author} {\bibfnamefont {D.}~\bibnamefont {Horsman}}, \bibinfo
  {author} {\bibfnamefont {M.}~\bibnamefont {Mhalla}},\ and\ \bibinfo {author}
  {\bibfnamefont {C.}~\bibnamefont {Branciard}},\ }\bibfield  {title} {\bibinfo
  {title} {Communication through coherent control of quantum channels},\
  }\href@noop {} {\bibfield  {journal} {\bibinfo  {journal} {Quantum}\ }\textbf
  {\bibinfo {volume} {4}},\ \bibinfo {pages} {333} (\bibinfo {year}
  {2020})}\BibitemShut {NoStop}%
\bibitem [{\citenamefont {Poggiali}\ \emph {et~al.}(2018)\citenamefont
  {Poggiali}, \citenamefont {Cappellaro},\ and\ \citenamefont
  {Fabbri}}]{poggiali2018optimal}%
  \BibitemOpen
  \bibfield  {author} {\bibinfo {author} {\bibfnamefont {F.}~\bibnamefont
  {Poggiali}}, \bibinfo {author} {\bibfnamefont {P.}~\bibnamefont
  {Cappellaro}},\ and\ \bibinfo {author} {\bibfnamefont {N.}~\bibnamefont
  {Fabbri}},\ }\bibfield  {title} {\bibinfo {title} {Optimal control for
  one-qubit quantum sensing},\ }\href@noop {} {\bibfield  {journal} {\bibinfo
  {journal} {Physical Review X}\ }\textbf {\bibinfo {volume} {8}},\ \bibinfo
  {pages} {021059} (\bibinfo {year} {2018})}\BibitemShut {NoStop}%
\bibitem [{\citenamefont {Soare}\ \emph {et~al.}(2014)\citenamefont {Soare},
  \citenamefont {Ball}, \citenamefont {Hayes}, \citenamefont {Sastrawan},
  \citenamefont {Jarratt}, \citenamefont {McLoughlin}, \citenamefont {Zhen},
  \citenamefont {Green},\ and\ \citenamefont
  {Biercuk}}]{soare2014experimental}%
  \BibitemOpen
  \bibfield  {author} {\bibinfo {author} {\bibfnamefont {A.}~\bibnamefont
  {Soare}}, \bibinfo {author} {\bibfnamefont {H.}~\bibnamefont {Ball}},
  \bibinfo {author} {\bibfnamefont {D.}~\bibnamefont {Hayes}}, \bibinfo
  {author} {\bibfnamefont {J.}~\bibnamefont {Sastrawan}}, \bibinfo {author}
  {\bibfnamefont {M.}~\bibnamefont {Jarratt}}, \bibinfo {author} {\bibfnamefont
  {J.}~\bibnamefont {McLoughlin}}, \bibinfo {author} {\bibfnamefont
  {X.}~\bibnamefont {Zhen}}, \bibinfo {author} {\bibfnamefont {T.}~\bibnamefont
  {Green}},\ and\ \bibinfo {author} {\bibfnamefont {M.}~\bibnamefont
  {Biercuk}},\ }\bibfield  {title} {\bibinfo {title} {Experimental noise
  filtering by quantum control},\ }\href@noop {} {\bibfield  {journal}
  {\bibinfo  {journal} {Nature Physics}\ }\textbf {\bibinfo {volume} {10}},\
  \bibinfo {pages} {825} (\bibinfo {year} {2014})}\BibitemShut {NoStop}%
\bibitem [{\citenamefont {Dong}\ and\ \citenamefont
  {Petersen}(2010)}]{dong2010quantum}%
  \BibitemOpen
  \bibfield  {author} {\bibinfo {author} {\bibfnamefont {D.}~\bibnamefont
  {Dong}}\ and\ \bibinfo {author} {\bibfnamefont {I.~R.}\ \bibnamefont
  {Petersen}},\ }\bibfield  {title} {\bibinfo {title} {Quantum control theory
  and applications: a survey},\ }\href@noop {} {\bibfield  {journal} {\bibinfo
  {journal} {IET control theory \& applications}\ }\textbf {\bibinfo {volume}
  {4}},\ \bibinfo {pages} {2651} (\bibinfo {year} {2010})}\BibitemShut
  {NoStop}%
\bibitem [{\citenamefont {Giannelli}\ \emph
  {et~al.}(2022{\natexlab{a}})\citenamefont {Giannelli}, \citenamefont {Sgroi},
  \citenamefont {Brown}, \citenamefont {Paraoanu}, \citenamefont {Paternostro},
  \citenamefont {Paladino},\ and\ \citenamefont
  {Falci}}]{giannelli2022tutorial}%
  \BibitemOpen
  \bibfield  {author} {\bibinfo {author} {\bibfnamefont {L.}~\bibnamefont
  {Giannelli}}, \bibinfo {author} {\bibfnamefont {S.}~\bibnamefont {Sgroi}},
  \bibinfo {author} {\bibfnamefont {J.}~\bibnamefont {Brown}}, \bibinfo
  {author} {\bibfnamefont {G.~S.}\ \bibnamefont {Paraoanu}}, \bibinfo {author}
  {\bibfnamefont {M.}~\bibnamefont {Paternostro}}, \bibinfo {author}
  {\bibfnamefont {E.}~\bibnamefont {Paladino}},\ and\ \bibinfo {author}
  {\bibfnamefont {G.}~\bibnamefont {Falci}},\ }\bibfield  {title} {\bibinfo
  {title} {A tutorial on optimal control and reinforcement learning methods for
  quantum technologies},\ }\href@noop {} {\bibfield  {journal} {\bibinfo
  {journal} {Physics Letters A}\ }\textbf {\bibinfo {volume} {434}},\ \bibinfo
  {pages} {128054} (\bibinfo {year} {2022}{\natexlab{a}})}\BibitemShut
  {NoStop}%
\bibitem [{\citenamefont {Mabuchi}(2008)}]{mabuchi2008coherent}%
  \BibitemOpen
  \bibfield  {author} {\bibinfo {author} {\bibfnamefont {H.}~\bibnamefont
  {Mabuchi}},\ }\bibfield  {title} {\bibinfo {title} {Coherent-feedback quantum
  control with a dynamic compensator},\ }\href@noop {} {\bibfield  {journal}
  {\bibinfo  {journal} {Physical Review A—Atomic, Molecular, and Optical
  Physics}\ }\textbf {\bibinfo {volume} {78}},\ \bibinfo {pages} {032323}
  (\bibinfo {year} {2008})}\BibitemShut {NoStop}%
\bibitem [{\citenamefont {Clark}\ \emph {et~al.}(2016)\citenamefont {Clark},
  \citenamefont {Stokes},\ and\ \citenamefont {Beige}}]{clark2016quantum}%
  \BibitemOpen
  \bibfield  {author} {\bibinfo {author} {\bibfnamefont {L.~A.}\ \bibnamefont
  {Clark}}, \bibinfo {author} {\bibfnamefont {A.}~\bibnamefont {Stokes}},\ and\
  \bibinfo {author} {\bibfnamefont {A.}~\bibnamefont {Beige}},\ }\bibfield
  {title} {\bibinfo {title} {Quantum-enhanced metrology with the single-mode
  coherent states of an optical cavity inside a quantum feedback loop},\
  }\href@noop {} {\bibfield  {journal} {\bibinfo  {journal} {Physical Review
  A}\ }\textbf {\bibinfo {volume} {94}},\ \bibinfo {pages} {023840} (\bibinfo
  {year} {2016})}\BibitemShut {NoStop}%
\bibitem [{\citenamefont {Karmakar}\ \emph {et~al.}(2025)\citenamefont
  {Karmakar}, \citenamefont {Lewalle}, \citenamefont {Zhang},\ and\
  \citenamefont {Whaley}}]{karmakar2025noise}%
  \BibitemOpen
  \bibfield  {author} {\bibinfo {author} {\bibfnamefont {T.}~\bibnamefont
  {Karmakar}}, \bibinfo {author} {\bibfnamefont {P.}~\bibnamefont {Lewalle}},
  \bibinfo {author} {\bibfnamefont {Y.}~\bibnamefont {Zhang}},\ and\ \bibinfo
  {author} {\bibfnamefont {K.~B.}\ \bibnamefont {Whaley}},\ }\bibfield  {title}
  {\bibinfo {title} {Noise-canceling quantum feedback: non-hermitian dynamics
  with applications to state preparation and magic state distillation},\
  }\href@noop {} {\bibfield  {journal} {\bibinfo  {journal} {arXiv preprint
  arXiv:2507.05611}\ } (\bibinfo {year} {2025})}\BibitemShut {NoStop}%
\bibitem [{\citenamefont {Viola}\ and\ \citenamefont {Lloyd}(1998)}]{DD1}%
  \BibitemOpen
  \bibfield  {author} {\bibinfo {author} {\bibfnamefont {L.}~\bibnamefont
  {Viola}}\ and\ \bibinfo {author} {\bibfnamefont {S.}~\bibnamefont {Lloyd}},\
  }\bibfield  {title} {\bibinfo {title} {Dynamical suppression of decoherence
  in two-state quantum systems},\ }\bibfield  {journal} {\bibinfo  {journal}
  {Physical Review A}\ }\href
  {https://doi.org/https://doi.org/10.1103/PhysRevA.58.2733}
  {https://doi.org/10.1103/PhysRevA.58.2733} (\bibinfo {year}
  {1998})\BibitemShut {NoStop}%
\bibitem [{\citenamefont {Carr}\ and\ \citenamefont {Purcell}(1954)}]{DD3}%
  \BibitemOpen
  \bibfield  {author} {\bibinfo {author} {\bibfnamefont {H.~Y.}\ \bibnamefont
  {Carr}}\ and\ \bibinfo {author} {\bibfnamefont {E.~M.}\ \bibnamefont
  {Purcell}},\ }\bibfield  {title} {\bibinfo {title} {Effects of diffusion on
  free precession in nuclear magnetic resonance experiments},\ }\bibfield
  {journal} {\bibinfo  {journal} {Physical Review}\ }\href
  {https://doi.org/https://doi.org/10.1103/PhysRev.94.630}
  {https://doi.org/10.1103/PhysRev.94.630} (\bibinfo {year} {1954})\BibitemShut
  {NoStop}%
\bibitem [{\citenamefont {Viola}\ \emph {et~al.}(1999)\citenamefont {Viola},
  \citenamefont {Knill},\ and\ \citenamefont {Lloyd}}]{DD5}%
  \BibitemOpen
  \bibfield  {author} {\bibinfo {author} {\bibfnamefont {L.}~\bibnamefont
  {Viola}}, \bibinfo {author} {\bibfnamefont {E.}~\bibnamefont {Knill}},\ and\
  \bibinfo {author} {\bibfnamefont {S.}~\bibnamefont {Lloyd}},\ }\bibfield
  {title} {\bibinfo {title} {Dynamical decoupling of open quantum systems},\
  }\bibfield  {journal} {\bibinfo  {journal} {Physical Review Letters}\ }\href
  {https://doi.org/https://doi.org/10.1103/PhysRevLett.82.2417}
  {https://doi.org/10.1103/PhysRevLett.82.2417} (\bibinfo {year}
  {1999})\BibitemShut {NoStop}%
\bibitem [{\citenamefont {Turyansky}\ \emph {et~al.}(2025)\citenamefont
  {Turyansky}, \citenamefont {Zolti}, \citenamefont {Cohen},\ and\
  \citenamefont {Pick}}]{turyansky2025pulse}%
  \BibitemOpen
  \bibfield  {author} {\bibinfo {author} {\bibfnamefont {D.}~\bibnamefont
  {Turyansky}}, \bibinfo {author} {\bibfnamefont {Y.}~\bibnamefont {Zolti}},
  \bibinfo {author} {\bibfnamefont {Y.}~\bibnamefont {Cohen}},\ and\ \bibinfo
  {author} {\bibfnamefont {A.}~\bibnamefont {Pick}},\ }\bibfield  {title}
  {\bibinfo {title} {Pulse optimization in adiabatic quantum computation and
  control},\ }\href@noop {} {\bibfield  {journal} {\bibinfo  {journal} {arXiv
  preprint arXiv:2507.09770}\ } (\bibinfo {year} {2025})}\BibitemShut {NoStop}%
\bibitem [{\citenamefont {Zeng}\ \emph {et~al.}(2019)\citenamefont {Zeng},
  \citenamefont {Gebremariam}, \citenamefont {Ding},\ and\ \citenamefont
  {Li}}]{zeng2019adiabatic}%
  \BibitemOpen
  \bibfield  {author} {\bibinfo {author} {\bibfnamefont {Y.-X.}\ \bibnamefont
  {Zeng}}, \bibinfo {author} {\bibfnamefont {T.}~\bibnamefont {Gebremariam}},
  \bibinfo {author} {\bibfnamefont {M.-S.}\ \bibnamefont {Ding}},\ and\
  \bibinfo {author} {\bibfnamefont {C.}~\bibnamefont {Li}},\ }\bibfield
  {title} {\bibinfo {title} {Adiabatic evolution: The influence of
  non-markovian characters on quantum adiabatic evolution (ann. phys.
  1/2019)},\ }\href@noop {} {\bibfield  {journal} {\bibinfo  {journal} {Annalen
  der Physik}\ }\textbf {\bibinfo {volume} {531}},\ \bibinfo {pages} {1970010}
  (\bibinfo {year} {2019})}\BibitemShut {NoStop}%
\bibitem [{\citenamefont {Mukherjee}\ \emph {et~al.}(2024)\citenamefont
  {Mukherjee}, \citenamefont {Penna}, \citenamefont {Cirinn{\`a}},
  \citenamefont {Paternostro}, \citenamefont {Paladino}, \citenamefont
  {Falci},\ and\ \citenamefont {Giannelli}}]{mukherjee2024noise}%
  \BibitemOpen
  \bibfield  {author} {\bibinfo {author} {\bibfnamefont {S.}~\bibnamefont
  {Mukherjee}}, \bibinfo {author} {\bibfnamefont {D.}~\bibnamefont {Penna}},
  \bibinfo {author} {\bibfnamefont {F.}~\bibnamefont {Cirinn{\`a}}}, \bibinfo
  {author} {\bibfnamefont {M.}~\bibnamefont {Paternostro}}, \bibinfo {author}
  {\bibfnamefont {E.}~\bibnamefont {Paladino}}, \bibinfo {author}
  {\bibfnamefont {G.}~\bibnamefont {Falci}},\ and\ \bibinfo {author}
  {\bibfnamefont {L.}~\bibnamefont {Giannelli}},\ }\bibfield  {title} {\bibinfo
  {title} {Noise classification in three-level quantum networks by machine
  learning},\ }\href@noop {} {\bibfield  {journal} {\bibinfo  {journal}
  {Machine Learning: Science and Technology}\ }\textbf {\bibinfo {volume}
  {5}},\ \bibinfo {pages} {045049} (\bibinfo {year} {2024})}\BibitemShut
  {NoStop}%
\bibitem [{\citenamefont {Delben}\ \emph {et~al.}(2023)\citenamefont {Delben},
  \citenamefont {Beims},\ and\ \citenamefont {da~Luz}}]{delben2023control}%
  \BibitemOpen
  \bibfield  {author} {\bibinfo {author} {\bibfnamefont {G.}~\bibnamefont
  {Delben}}, \bibinfo {author} {\bibfnamefont {M.}~\bibnamefont {Beims}},\ and\
  \bibinfo {author} {\bibfnamefont {M.}~\bibnamefont {da~Luz}},\ }\bibfield
  {title} {\bibinfo {title} {Control of a qubit under markovian and
  non-markovian noise},\ }\href@noop {} {\bibfield  {journal} {\bibinfo
  {journal} {Physical Review A}\ }\textbf {\bibinfo {volume} {108}},\ \bibinfo
  {pages} {012620} (\bibinfo {year} {2023})}\BibitemShut {NoStop}%
\bibitem [{\citenamefont {Ortega-Taberner}\ \emph {et~al.}(2024)\citenamefont
  {Ortega-Taberner}, \citenamefont {O’Neill}, \citenamefont {Butler},
  \citenamefont {Fux},\ and\ \citenamefont {Eastham}}]{ortega2024unifying}%
  \BibitemOpen
  \bibfield  {author} {\bibinfo {author} {\bibfnamefont {C.}~\bibnamefont
  {Ortega-Taberner}}, \bibinfo {author} {\bibfnamefont {E.}~\bibnamefont
  {O’Neill}}, \bibinfo {author} {\bibfnamefont {E.}~\bibnamefont {Butler}},
  \bibinfo {author} {\bibfnamefont {G.~E.}\ \bibnamefont {Fux}},\ and\ \bibinfo
  {author} {\bibfnamefont {P.}~\bibnamefont {Eastham}},\ }\bibfield  {title}
  {\bibinfo {title} {Unifying methods for optimal control in non-markovian
  quantum systems via process tensors},\ }\href@noop {} {\bibfield  {journal}
  {\bibinfo  {journal} {The Journal of Chemical Physics}\ }\textbf {\bibinfo
  {volume} {161}} (\bibinfo {year} {2024})}\BibitemShut {NoStop}%
\bibitem [{\citenamefont {Youssry}\ \emph {et~al.}(2020)\citenamefont
  {Youssry}, \citenamefont {Paz-Silva},\ and\ \citenamefont
  {Ferrie}}]{youssri1}%
  \BibitemOpen
  \bibfield  {author} {\bibinfo {author} {\bibfnamefont {A.}~\bibnamefont
  {Youssry}}, \bibinfo {author} {\bibfnamefont {G.~A.}\ \bibnamefont
  {Paz-Silva}},\ and\ \bibinfo {author} {\bibfnamefont {C.}~\bibnamefont
  {Ferrie}},\ }\bibfield  {title} {\bibinfo {title} {Characterization and
  control of open quantum systems beyond quantum noise spectroscopy},\
  }\bibfield  {journal} {\bibinfo  {journal} {npj (Nature Partner Journals)
  Quantum Information}\ }\href
  {https://doi.org/https://doi.org/10.1038/s41534-020-00332-8}
  {https://doi.org/10.1038/s41534-020-00332-8} (\bibinfo {year}
  {2020})\BibitemShut {NoStop}%
\bibitem [{\citenamefont {Auza}\ \emph {et~al.}()\citenamefont {Auza},
  \citenamefont {Youssry}, \citenamefont {Paz-Silva},\ and\ \citenamefont
  {Peruzzo}}]{auza}%
  \BibitemOpen
  \bibfield  {author} {\bibinfo {author} {\bibfnamefont {A.}~\bibnamefont
  {Auza}}, \bibinfo {author} {\bibfnamefont {A.}~\bibnamefont {Youssry}},
  \bibinfo {author} {\bibfnamefont {G.}~\bibnamefont {Paz-Silva}},\ and\
  \bibinfo {author} {\bibfnamefont {A.}~\bibnamefont {Peruzzo}},\ }\bibfield
  {title} {\bibinfo {title} {Quantum control in the presence of strongly
  coupled non-markovian noise},\ }\bibfield  {journal} {\bibinfo  {journal}
  {arXiv}\ }\href {https://doi.org/https://doi.org/10.48550/arXiv.2404.19251}
  {https://doi.org/10.48550/arXiv.2404.19251}\BibitemShut {NoStop}%
\bibitem [{\citenamefont {Youssry}\ and\ \citenamefont
  {Nurdin}(2023)}]{youssri2}%
  \BibitemOpen
  \bibfield  {author} {\bibinfo {author} {\bibfnamefont {A.}~\bibnamefont
  {Youssry}}\ and\ \bibinfo {author} {\bibfnamefont {H.~I.}\ \bibnamefont
  {Nurdin}},\ }\bibfield  {title} {\bibinfo {title} {Multi-axis control of a
  qubit in the presence of unknown non-markovian quantum noise},\ }\bibfield
  {journal} {\bibinfo  {journal} {Quantum Sci. Technol.}\ }\href
  {https://doi.org/https://doi.org/10.1088/2058-9565/aca711}
  {https://doi.org/10.1088/2058-9565/aca711} (\bibinfo {year}
  {2023})\BibitemShut {NoStop}%
\bibitem [{\citenamefont {Youssry}\ \emph {et~al.}(2024)\citenamefont
  {Youssry}, \citenamefont {Yang},\ and\ \citenamefont {et~al.}}]{youssri3}%
  \BibitemOpen
  \bibfield  {author} {\bibinfo {author} {\bibfnamefont {A.}~\bibnamefont
  {Youssry}}, \bibinfo {author} {\bibfnamefont {Y.}~\bibnamefont {Yang}},\ and\
  \bibinfo {author} {\bibfnamefont {R.~J.~C.}\ \bibnamefont {et~al.}},\
  }\bibfield  {title} {\bibinfo {title} {Experimental graybox quantum system
  identification and control},\ }\bibfield  {journal} {\bibinfo  {journal} {npj
  Quantum Information}\ }\href {https://doi.org/https://doi.org/10.1038/}
  {https://doi.org/10.1038/} (\bibinfo {year} {2024})\BibitemShut {NoStop}%
\bibitem [{\citenamefont {Gurney}(2018)}]{gurney2018introduction}%
  \BibitemOpen
  \bibfield  {author} {\bibinfo {author} {\bibfnamefont {K.}~\bibnamefont
  {Gurney}},\ }\href@noop {} {\emph {\bibinfo {title} {An introduction to
  neural networks}}}\ (\bibinfo  {publisher} {CRC press},\ \bibinfo {year}
  {2018})\BibitemShut {NoStop}%
\bibitem [{\citenamefont {Marquardt}(2021)}]{marquardt2021machine}%
  \BibitemOpen
  \bibfield  {author} {\bibinfo {author} {\bibfnamefont {F.}~\bibnamefont
  {Marquardt}},\ }\bibfield  {title} {\bibinfo {title} {Machine learning and
  quantum devices},\ }\href@noop {} {\bibfield  {journal} {\bibinfo  {journal}
  {SciPost Physics Lecture Notes}\ ,\ \bibinfo {pages} {029}} (\bibinfo {year}
  {2021})}\BibitemShut {NoStop}%
\bibitem [{\citenamefont {G{\'e}ron}(2022)}]{geron2022hands}%
  \BibitemOpen
  \bibfield  {author} {\bibinfo {author} {\bibfnamefont {A.}~\bibnamefont
  {G{\'e}ron}},\ }\href@noop {} {\emph {\bibinfo {title} {Hands-on machine
  learning with Scikit-Learn, Keras, and TensorFlow}}}\ (\bibinfo  {publisher}
  {" O'Reilly Media, Inc."},\ \bibinfo {year} {2022})\BibitemShut {NoStop}%
\bibitem [{\citenamefont {Nasteski}(2017)}]{nasteski2017overview}%
  \BibitemOpen
  \bibfield  {author} {\bibinfo {author} {\bibfnamefont {V.}~\bibnamefont
  {Nasteski}},\ }\bibfield  {title} {\bibinfo {title} {An overview of the
  supervised machine learning methods},\ }\href@noop {} {\bibfield  {journal}
  {\bibinfo  {journal} {Horizons. b}\ }\textbf {\bibinfo {volume} {4}},\
  \bibinfo {pages} {56} (\bibinfo {year} {2017})}\BibitemShut {NoStop}%
\bibitem [{\citenamefont {Perrier}\ \emph {et~al.}(2022)\citenamefont
  {Perrier}, \citenamefont {Youssri},\ and\ \citenamefont {Ferrie}}]{qdataset}%
  \BibitemOpen
  \bibfield  {author} {\bibinfo {author} {\bibfnamefont {E.}~\bibnamefont
  {Perrier}}, \bibinfo {author} {\bibfnamefont {A.}~\bibnamefont {Youssri}},\
  and\ \bibinfo {author} {\bibfnamefont {C.}~\bibnamefont {Ferrie}},\
  }\bibfield  {title} {\bibinfo {title} {Qdataset, quantum datasets for machine
  learning},\ }\bibfield  {journal} {\bibinfo  {journal} {Nature - scientific
  data}\ }\href {https://doi.org/https://doi.org/10.1038/s41597-022-01639-1}
  {https://doi.org/10.1038/s41597-022-01639-1} (\bibinfo {year}
  {2022})\BibitemShut {NoStop}%
\bibitem [{\citenamefont {Gardiner}(1983)}]{gardiner}%
  \BibitemOpen
  \bibfield  {author} {\bibinfo {author} {\bibfnamefont {C.~W.}\ \bibnamefont
  {Gardiner}},\ }\href@noop {} {\emph {\bibinfo {title} {Handbook of Stochastic
  Methods - For Physics, Chemistry and the Natural Sciences}}}\ (\bibinfo
  {publisher} {Springer},\ \bibinfo {year} {1983})\BibitemShut {NoStop}%
\bibitem [{\citenamefont {Mandel}\ and\ \citenamefont
  {Wolf}(1995)}]{mandel-wolf}%
  \BibitemOpen
  \bibfield  {author} {\bibinfo {author} {\bibfnamefont {L.}~\bibnamefont
  {Mandel}}\ and\ \bibinfo {author} {\bibfnamefont {E.}~\bibnamefont {Wolf}},\
  }\href@noop {} {\emph {\bibinfo {title} {Optical coherence and quantum
  optics}}}\ (\bibinfo  {publisher} {Cambridge University Press},\ \bibinfo
  {year} {1995})\BibitemShut {NoStop}%
\bibitem [{\citenamefont {Papoulis}\ and\ \citenamefont
  {Pillai}(2002)}]{papoulis}%
  \BibitemOpen
  \bibfield  {author} {\bibinfo {author} {\bibfnamefont {A.}~\bibnamefont
  {Papoulis}}\ and\ \bibinfo {author} {\bibfnamefont {S.~U.}\ \bibnamefont
  {Pillai}},\ }\href@noop {} {\emph {\bibinfo {title} {Probability, Random
  Variables and Stochastic Processes}}}\ (\bibinfo  {publisher} {McGraw-Hill},\
  \bibinfo {year} {2002})\BibitemShut {NoStop}%
\bibitem [{\citenamefont {Uhlenbeck}\ and\ \citenamefont
  {Ornstein}(1930)}]{OU1}%
  \BibitemOpen
  \bibfield  {author} {\bibinfo {author} {\bibfnamefont {G.~E.}\ \bibnamefont
  {Uhlenbeck}}\ and\ \bibinfo {author} {\bibfnamefont {L.~S.}\ \bibnamefont
  {Ornstein}},\ }\bibfield  {title} {\bibinfo {title} {On the theory of the
  brownian motion},\ }\bibfield  {journal} {\bibinfo  {journal} {Physical
  Review}\ }\href {https://doi.org/https://doi.org/10.1103/PhysRev.36.823}
  {https://doi.org/10.1103/PhysRev.36.823} (\bibinfo {year} {1930})\BibitemShut
  {NoStop}%
\bibitem [{\citenamefont {Gillespie}(1996)}]{OU2}%
  \BibitemOpen
  \bibfield  {author} {\bibinfo {author} {\bibfnamefont {D.~T.}\ \bibnamefont
  {Gillespie}},\ }\bibfield  {title} {\bibinfo {title} {Exact numerical
  simulation of the ornstein-uhlenbeck process and its integral},\ }\bibfield
  {journal} {\bibinfo  {journal} {Physical Review E}\ }\href
  {https://doi.org/https://doi.org/10.1103/PhysRevE.54.2084}
  {https://doi.org/10.1103/PhysRevE.54.2084} (\bibinfo {year}
  {1996})\BibitemShut {NoStop}%
\bibitem [{\citenamefont {Paladino}\ \emph
  {et~al.}(2002{\natexlab{a}})\citenamefont {Paladino}, \citenamefont {Faoro},
  \citenamefont {D'Arrigo},\ and\ \citenamefont {Falci}}]{falci1}%
  \BibitemOpen
  \bibfield  {author} {\bibinfo {author} {\bibfnamefont {E.}~\bibnamefont
  {Paladino}}, \bibinfo {author} {\bibfnamefont {L.}~\bibnamefont {Faoro}},
  \bibinfo {author} {\bibfnamefont {A.}~\bibnamefont {D'Arrigo}},\ and\
  \bibinfo {author} {\bibfnamefont {G.}~\bibnamefont {Falci}},\ }\bibfield
  {title} {\bibinfo {title} {Background charges induced stochastic fluctuations
  in josephson qubits},\ }\bibfield  {journal} {\bibinfo  {journal} {Journal of
  the Physical Society of Japan}\ }\href
  {https://doi.org/https://doi.org/10.1143/JPSJS.72SA.165}
  {https://doi.org/10.1143/JPSJS.72SA.165} (\bibinfo {year}
  {2002}{\natexlab{a}})\BibitemShut {NoStop}%
\bibitem [{\citenamefont {Paladino}\ \emph {et~al.}(2003)\citenamefont
  {Paladino}, \citenamefont {Faoro},\ and\ \citenamefont {Falci}}]{falci2}%
  \BibitemOpen
  \bibfield  {author} {\bibinfo {author} {\bibfnamefont {E.}~\bibnamefont
  {Paladino}}, \bibinfo {author} {\bibfnamefont {L.}~\bibnamefont {Faoro}},\
  and\ \bibinfo {author} {\bibfnamefont {G.}~\bibnamefont {Falci}},\ }\bibfield
   {title} {\bibinfo {title} {Decoherence due to discrete noise in josephson
  qubits},\ }\bibfield  {journal} {\bibinfo  {journal} {Chapter in Advances in
  Solid State Physics (ASSP, volume 43)}\ }\href
  {https://doi.org/https://doi.org/10.1007/978-3-540-44838-9\_53}
  {https://doi.org/10.1007/978-3-540-44838-9\_53} (\bibinfo {year}
  {2003})\BibitemShut {NoStop}%
\bibitem [{\citenamefont {Paladino}\ \emph
  {et~al.}(2002{\natexlab{b}})\citenamefont {Paladino}, \citenamefont {Faoro},
  \citenamefont {Falci},\ and\ \citenamefont {Fazio}}]{paladino2}%
  \BibitemOpen
  \bibfield  {author} {\bibinfo {author} {\bibfnamefont {E.}~\bibnamefont
  {Paladino}}, \bibinfo {author} {\bibfnamefont {L.}~\bibnamefont {Faoro}},
  \bibinfo {author} {\bibfnamefont {G.}~\bibnamefont {Falci}},\ and\ \bibinfo
  {author} {\bibfnamefont {R.}~\bibnamefont {Fazio}},\ }\bibfield  {title}
  {\bibinfo {title} {Decoherence and 1/f noise in josephson qubits},\
  }\bibfield  {journal} {\bibinfo  {journal} {Physical Review Letters}\ }\href
  {https://doi.org/https://doi.org/10.1103/PhysRevLett.88.228304}
  {https://doi.org/10.1103/PhysRevLett.88.228304} (\bibinfo {year}
  {2002}{\natexlab{b}})\BibitemShut {NoStop}%
\bibitem [{\citenamefont {Ripoll}(2022)}]{supercondqubit}%
  \BibitemOpen
  \bibfield  {author} {\bibinfo {author} {\bibfnamefont {J.~J.~G.}\
  \bibnamefont {Ripoll}},\ }\href@noop {} {\emph {\bibinfo {title} {Quantum
  Information and Quantum Optics with Superconducting Circuits}}}\ (\bibinfo
  {publisher} {Cambridge University Press},\ \bibinfo {year}
  {2022})\BibitemShut {NoStop}%
\bibitem [{\citenamefont {D'Arrigo}\ \emph {et~al.}(2024)\citenamefont
  {D'Arrigo}, \citenamefont {Piccitto}, \citenamefont {Falci},\ and\
  \citenamefont {Paladino}}]{falci3}%
  \BibitemOpen
  \bibfield  {author} {\bibinfo {author} {\bibfnamefont {A.}~\bibnamefont
  {D'Arrigo}}, \bibinfo {author} {\bibfnamefont {G.}~\bibnamefont {Piccitto}},
  \bibinfo {author} {\bibfnamefont {G.}~\bibnamefont {Falci}},\ and\ \bibinfo
  {author} {\bibfnamefont {E.}~\bibnamefont {Paladino}},\ }\bibfield  {title}
  {\bibinfo {title} {Open‑loop quantum control of small‑size networks for
  high‑order cumulants and cross‑correlations sensing},\ }\bibfield
  {journal} {\bibinfo  {journal} {Nature Scientific Reports}\ }\href
  {https://doi.org/https://doi.org/10.1038/s41598-024-67503-x}
  {https://doi.org/10.1038/s41598-024-67503-x} (\bibinfo {year}
  {2024})\BibitemShut {NoStop}%
\bibitem [{\citenamefont {Wood}\ \emph {et~al.}(2015)\citenamefont {Wood},
  \citenamefont {Biamonte},\ and\ \citenamefont {Cory}}]{tensornet}%
  \BibitemOpen
  \bibfield  {author} {\bibinfo {author} {\bibfnamefont {C.~J.}\ \bibnamefont
  {Wood}}, \bibinfo {author} {\bibfnamefont {J.~D.}\ \bibnamefont {Biamonte}},\
  and\ \bibinfo {author} {\bibfnamefont {D.~G.}\ \bibnamefont {Cory}},\
  }\bibfield  {title} {\bibinfo {title} {Tensor networks and graphical calculus
  for open quantum systems}\ }\href
  {https://doi.org/https://doi.org/10.48550/arXiv.1111.6950}
  {https://doi.org/10.48550/arXiv.1111.6950} (\bibinfo {year}
  {2015})\BibitemShut {NoStop}%
\bibitem [{\citenamefont {Giannelli}\ \emph
  {et~al.}(2022{\natexlab{b}})\citenamefont {Giannelli}, \citenamefont {Sgroi},
  \citenamefont {Brown}, \citenamefont {Paraoanu}, \citenamefont {Paternostro},
  \citenamefont {Paladino},\ and\ \citenamefont {Falci}}]{giannelli}%
  \BibitemOpen
  \bibfield  {author} {\bibinfo {author} {\bibfnamefont {L.}~\bibnamefont
  {Giannelli}}, \bibinfo {author} {\bibfnamefont {P.}~\bibnamefont {Sgroi}},
  \bibinfo {author} {\bibfnamefont {J.}~\bibnamefont {Brown}}, \bibinfo
  {author} {\bibfnamefont {G.~S.}\ \bibnamefont {Paraoanu}}, \bibinfo {author}
  {\bibfnamefont {M.}~\bibnamefont {Paternostro}}, \bibinfo {author}
  {\bibfnamefont {E.}~\bibnamefont {Paladino}},\ and\ \bibinfo {author}
  {\bibfnamefont {G.}~\bibnamefont {Falci}},\ }\bibfield  {title} {\bibinfo
  {title} {A tutorial on optimal control and reinforcement learning methods for
  quantum technologies},\ }\bibfield  {journal} {\bibinfo  {journal} {Physics
  Letters A}\ }\href
  {https://doi.org/https://doi.org/10.1016/j.physleta.2022.128054}
  {https://doi.org/10.1016/j.physleta.2022.128054} (\bibinfo {year}
  {2022}{\natexlab{b}})\BibitemShut {NoStop}%
\bibitem [{\citenamefont {Wilhelm}\ \emph {et~al.}(2020)\citenamefont
  {Wilhelm}, \citenamefont {Kirchhoff}, \citenamefont {Machnes}, \citenamefont
  {Wittler},\ and\ \citenamefont {Sugny}}]{wilhelm}%
  \BibitemOpen
  \bibfield  {author} {\bibinfo {author} {\bibfnamefont {F.~K.}\ \bibnamefont
  {Wilhelm}}, \bibinfo {author} {\bibfnamefont {S.}~\bibnamefont {Kirchhoff}},
  \bibinfo {author} {\bibfnamefont {S.}~\bibnamefont {Machnes}}, \bibinfo
  {author} {\bibfnamefont {N.}~\bibnamefont {Wittler}},\ and\ \bibinfo {author}
  {\bibfnamefont {D.}~\bibnamefont {Sugny}},\ }\bibfield  {title} {\bibinfo
  {title} {An introduction into optimal control for quantum technologies},\
  }\href {https://arxiv.org/pdf/2003.10132} {\bibfield  {journal} {\bibinfo
  {journal} {arXiv:2003.10132 [quant-ph]}\ } (\bibinfo {year}
  {2020})}\BibitemShut {NoStop}%
\bibitem [{\citenamefont {Liberzon}(2012)}]{oc2}%
  \BibitemOpen
  \bibfield  {author} {\bibinfo {author} {\bibfnamefont {D.}~\bibnamefont
  {Liberzon}},\ }\href@noop {} {\emph {\bibinfo {title} {Calculus of Variations
  and Optimal Control Theory: A Concise Introduction}}}\ (\bibinfo  {publisher}
  {Princeton University Press},\ \bibinfo {year} {2012})\BibitemShut {NoStop}%
\bibitem [{\citenamefont {Nocedal}\ and\ \citenamefont
  {Wright}(2006)}]{optimizers}%
  \BibitemOpen
  \bibfield  {author} {\bibinfo {author} {\bibfnamefont {J.}~\bibnamefont
  {Nocedal}}\ and\ \bibinfo {author} {\bibfnamefont {S.~J.}\ \bibnamefont
  {Wright}},\ }\href@noop {} {\emph {\bibinfo {title} {Numerical
  Optimization}}}\ (\bibinfo  {publisher} {Springer},\ \bibinfo {year}
  {2006})\BibitemShut {NoStop}%
\bibitem [{\citenamefont {Vaswani}\ \emph {et~al.}(2017)\citenamefont
  {Vaswani}, \citenamefont {Shazeer}, \citenamefont {Parmar}, \citenamefont
  {Uszkoreit}, \citenamefont {Jones}, \citenamefont {Gomez}, \citenamefont
  {Łukasz Kaiser},\ and\ \citenamefont {Polosukhin}}]{attention}%
  \BibitemOpen
  \bibfield  {author} {\bibinfo {author} {\bibfnamefont {A.}~\bibnamefont
  {Vaswani}}, \bibinfo {author} {\bibfnamefont {N.}~\bibnamefont {Shazeer}},
  \bibinfo {author} {\bibfnamefont {N.}~\bibnamefont {Parmar}}, \bibinfo
  {author} {\bibfnamefont {J.}~\bibnamefont {Uszkoreit}}, \bibinfo {author}
  {\bibfnamefont {L.}~\bibnamefont {Jones}}, \bibinfo {author} {\bibfnamefont
  {A.~N.}\ \bibnamefont {Gomez}}, \bibinfo {author} {\bibnamefont {Łukasz
  Kaiser}},\ and\ \bibinfo {author} {\bibfnamefont {I.}~\bibnamefont
  {Polosukhin}},\ }\bibfield  {title} {\bibinfo {title} {Attention is all you
  need},\ }\href@noop {} {\bibfield  {journal} {\bibinfo  {journal}
  {Proceedings of the 31st Conference on Neural Information Processing Systems
  (NIPS 2017)}\ } (\bibinfo {year} {2017})}\BibitemShut {NoStop}%
\bibitem [{\citenamefont {Kingma}\ and\ \citenamefont {Ba}(2015)}]{adam}%
  \BibitemOpen
  \bibfield  {author} {\bibinfo {author} {\bibfnamefont {D.~P.}\ \bibnamefont
  {Kingma}}\ and\ \bibinfo {author} {\bibfnamefont {J.}~\bibnamefont {Ba}},\
  }\bibfield  {title} {\bibinfo {title} {Adam: A method for stochastic
  optimization},\ }\href@noop {} {\bibfield  {journal} {\bibinfo  {journal}
  {3rd International Conference on Learning Representations, Conference Track
  Proceedings}\ } (\bibinfo {year} {2015})}\BibitemShut {NoStop}%
\bibitem [{\citenamefont {Saha}(2015)}]{python}%
  \BibitemOpen
  \bibfield  {author} {\bibinfo {author} {\bibfnamefont {A.}~\bibnamefont
  {Saha}},\ }\href@noop {} {\emph {\bibinfo {title} {Doing Math with Python}}}\
  (\bibinfo  {publisher} {No Starch Press, Inc.},\ \bibinfo {year}
  {2015})\BibitemShut {NoStop}%
\bibitem [{\citenamefont {et~al.}(2015{\natexlab{a}})}]{tensorflow}%
  \BibitemOpen
  \bibfield  {author} {\bibinfo {author} {\bibfnamefont {M.~A.}\ \bibnamefont
  {et~al.}},\ }\bibfield  {title} {\bibinfo {title} {Tensorflow: Large-scale
  machine learning on heterogeneous systems},\ }\href
  {https://www.tensorflow.org/} {\  (\bibinfo {year}
  {2015}{\natexlab{a}})}\BibitemShut {NoStop}%
\bibitem [{\citenamefont {et~al.}(2015{\natexlab{b}})}]{keras}%
  \BibitemOpen
  \bibfield  {author} {\bibinfo {author} {\bibfnamefont {F.~C.}\ \bibnamefont
  {et~al.}},\ }\bibfield  {title} {\bibinfo {title} {Keras},\ }\href
  {https://keras.io} {\  (\bibinfo {year} {2015}{\natexlab{b}})}\BibitemShut
  {NoStop}%
\bibitem [{\citenamefont {Paz-Silva}\ \emph {et~al.}(2019)\citenamefont
  {Paz-Silva}, \citenamefont {Norris}, \citenamefont {Beaudoin},\ and\
  \citenamefont {Viola}}]{paz-silva}%
  \BibitemOpen
  \bibfield  {author} {\bibinfo {author} {\bibfnamefont {G.~A.}\ \bibnamefont
  {Paz-Silva}}, \bibinfo {author} {\bibfnamefont {L.~M.}\ \bibnamefont
  {Norris}}, \bibinfo {author} {\bibfnamefont {F.}~\bibnamefont {Beaudoin}},\
  and\ \bibinfo {author} {\bibfnamefont {L.}~\bibnamefont {Viola}},\ }\bibfield
   {title} {\bibinfo {title} {Extending comb-based spectral estimation to
  multiaxis quantum noise},\ }\bibfield  {journal} {\bibinfo  {journal} {Phys.
  Rev. Lett. A}\ }\href
  {https://doi.org/https://doi.org/10.1103/PhysRevA.100.042334}
  {https://doi.org/10.1103/PhysRevA.100.042334} (\bibinfo {year}
  {2019})\BibitemShut {NoStop}%
\bibitem [{\citenamefont {Shen}\ \emph {et~al.}(2021)\citenamefont {Shen},
  \citenamefont {Zhang}, \citenamefont {Zhao}, \citenamefont {Yi},\ and\
  \citenamefont {Li}}]{effattention}%
  \BibitemOpen
  \bibfield  {author} {\bibinfo {author} {\bibfnamefont {Z.}~\bibnamefont
  {Shen}}, \bibinfo {author} {\bibfnamefont {M.}~\bibnamefont {Zhang}},
  \bibinfo {author} {\bibfnamefont {H.}~\bibnamefont {Zhao}}, \bibinfo {author}
  {\bibfnamefont {S.}~\bibnamefont {Yi}},\ and\ \bibinfo {author}
  {\bibfnamefont {H.}~\bibnamefont {Li}},\ }\bibfield  {title} {\bibinfo
  {title} {Efficient attention: Attention with linear complexities},\
  }\bibfield  {journal} {\bibinfo  {journal} {2021 IEEE Winter Conference on
  Applications of Computer Vision (WACV)}\ }\href
  {https://doi.org/10.1109/WACV48630.2021.00357} {10.1109/WACV48630.2021.00357}
  (\bibinfo {year} {2021})\BibitemShut {NoStop}%
\bibitem [{\citenamefont {He}\ \emph {et~al.}(2016)\citenamefont {He},
  \citenamefont {Zhang}, \citenamefont {Ren},\ and\ \citenamefont
  {Sun}}]{he2016deep}%
  \BibitemOpen
  \bibfield  {author} {\bibinfo {author} {\bibfnamefont {K.}~\bibnamefont
  {He}}, \bibinfo {author} {\bibfnamefont {X.}~\bibnamefont {Zhang}}, \bibinfo
  {author} {\bibfnamefont {S.}~\bibnamefont {Ren}},\ and\ \bibinfo {author}
  {\bibfnamefont {J.}~\bibnamefont {Sun}},\ }\bibfield  {title} {\bibinfo
  {title} {Deep residual learning for image recognition},\ }in\ \href@noop {}
  {\emph {\bibinfo {booktitle} {Proceedings of the IEEE Conference on Computer
  Vision and Pattern Recognition (CVPR)}}}\ (\bibinfo {year} {2016})\ pp.\
  \bibinfo {pages} {770--778}\BibitemShut {NoStop}%
\bibitem [{\citenamefont {Ba}\ \emph {et~al.}(2016)\citenamefont {Ba},
  \citenamefont {Kiros},\ and\ \citenamefont {Hinton}}]{ba2016layer}%
  \BibitemOpen
  \bibfield  {author} {\bibinfo {author} {\bibfnamefont {J.~L.}\ \bibnamefont
  {Ba}}, \bibinfo {author} {\bibfnamefont {J.~R.}\ \bibnamefont {Kiros}},\ and\
  \bibinfo {author} {\bibfnamefont {G.~E.}\ \bibnamefont {Hinton}},\ }\bibfield
   {title} {\bibinfo {title} {Layer normalization},\ }\href@noop {} {\bibfield
  {journal} {\bibinfo  {journal} {arXiv preprint arXiv:1607.06450}\ } (\bibinfo
  {year} {2016})}\BibitemShut {NoStop}%
\bibitem [{\citenamefont {Hendrycks}\ and\ \citenamefont
  {Gimpel}(2016)}]{hendrycks2016gaussian}%
  \BibitemOpen
  \bibfield  {author} {\bibinfo {author} {\bibfnamefont {D.}~\bibnamefont
  {Hendrycks}}\ and\ \bibinfo {author} {\bibfnamefont {K.}~\bibnamefont
  {Gimpel}},\ }\bibfield  {title} {\bibinfo {title} {Gaussian error linear
  units (gelus)},\ }\href@noop {} {\bibfield  {journal} {\bibinfo  {journal}
  {arXiv preprint arXiv:1606.08415}\ } (\bibinfo {year} {2016})}\BibitemShut
  {NoStop}%
\bibitem [{\citenamefont {Lin}\ \emph {et~al.}(2014)\citenamefont {Lin},
  \citenamefont {Chen},\ and\ \citenamefont {Yan}}]{lin2014network}%
  \BibitemOpen
  \bibfield  {author} {\bibinfo {author} {\bibfnamefont {M.}~\bibnamefont
  {Lin}}, \bibinfo {author} {\bibfnamefont {Q.}~\bibnamefont {Chen}},\ and\
  \bibinfo {author} {\bibfnamefont {S.}~\bibnamefont {Yan}},\ }\bibfield
  {title} {\bibinfo {title} {Network in network},\ }in\ \href@noop {} {\emph
  {\bibinfo {booktitle} {International Conference on Learning Representations
  (ICLR)}}}\ (\bibinfo {year} {2014})\BibitemShut {NoStop}%
\bibitem [{\citenamefont {Srivastava}\ \emph {et~al.}(2014)\citenamefont
  {Srivastava}, \citenamefont {Hinton}, \citenamefont {Krizhevsky},
  \citenamefont {Sutskever},\ and\ \citenamefont
  {Salakhutdinov}}]{srivastava2014dropout}%
  \BibitemOpen
  \bibfield  {author} {\bibinfo {author} {\bibfnamefont {N.}~\bibnamefont
  {Srivastava}}, \bibinfo {author} {\bibfnamefont {G.}~\bibnamefont {Hinton}},
  \bibinfo {author} {\bibfnamefont {A.}~\bibnamefont {Krizhevsky}}, \bibinfo
  {author} {\bibfnamefont {I.}~\bibnamefont {Sutskever}},\ and\ \bibinfo
  {author} {\bibfnamefont {R.}~\bibnamefont {Salakhutdinov}},\ }\bibfield
  {title} {\bibinfo {title} {Dropout: A simple way to prevent neural networks
  from overfitting},\ }\href@noop {} {\bibfield  {journal} {\bibinfo  {journal}
  {Journal of Machine Learning Research}\ }\textbf {\bibinfo {volume} {15}},\
  \bibinfo {pages} {1929} (\bibinfo {year} {2014})}\BibitemShut {NoStop}%
\bibitem [{\citenamefont {Loshchilov}\ and\ \citenamefont
  {Hutter}(2017)}]{loshchilov2017sgdr}%
  \BibitemOpen
  \bibfield  {author} {\bibinfo {author} {\bibfnamefont {I.}~\bibnamefont
  {Loshchilov}}\ and\ \bibinfo {author} {\bibfnamefont {F.}~\bibnamefont
  {Hutter}},\ }\bibfield  {title} {\bibinfo {title} {Sgdr: Stochastic gradient
  descent with warm restarts},\ }in\ \href@noop {} {\emph {\bibinfo {booktitle}
  {International Conference on Learning Representations (ICLR)}}}\ (\bibinfo
  {year} {2017})\BibitemShut {NoStop}%
\end{thebibliography}%
\end{document}